\documentclass[sigplan,nonacm]{acmart}
\setcopyright{none}
\AtBeginDocument{%
  }

\renewcommand\footnotetextcopyrightpermission[1]{}
\usepackage{tikz}
\usepackage{amsmath}

\usepackage{filecontents}

\usepackage{algorithm} 
\usepackage{amsfonts}
\usepackage{booktabs}
\usepackage{nameref}
\usepackage{framed}
\usepackage[nounderscore]{syntax}
\usepackage{algpseudocode} 
\usepackage{ragged2e}
\usepackage{simplebnf}
\usepackage{mathpartir}
\usepackage{stmaryrd}
\usepackage{tabularx}
\usepackage{xspace}
\usepackage{enumitem}
    \makeatletter
    \algrenewcommand\ALG@beginalgorithmic{\footnotesize}
    
    \makeatother
\usepackage{listings}
\usepackage{xcolor}

\lstdefinelanguage{CustomC}{
  language=C,
  morekeywords=[2]{bpf_map_lookup_elem,bpf_map_update_elem, xdp_fw_prog, xdp_fw, bpf_redirect_map},
  morecomment=[l]{//},
  sensitive=true,
  moredelim=**[is][\bfseries\color{red!80!black}]{@}{@},
  moredelim=**[is][\footnotesize\bfseries\color{orange!90!red}]{`}{`},
  moredelim=**[is][\footnotesize\bfseries\color{blue!60!black}]{?}{?}
}

\usepackage{tikz}
\usepackage{amsmath}
\usepackage{multirow}
\usetikzlibrary{calc}
\usetikzlibrary{patterns}
\usepackage[normalem]{ulem}

\usepackage{pgfplots}
\usepackage{pgfplotstable}
\pgfplotsset{compat=1.17}

\usepackage{filecontents}
\usepackage[font=small,skip=1.5pt]{caption}
\usepackage{subcaption}
\AtBeginEnvironment{grammar}{\footnotesize}

\usepackage{pifont}

\begin{document}

% New definitions
\algnewcommand\algorithmicswitch{\textbf{switch}}
\algnewcommand\algorithmiccase{\textbf{case}}
\algnewcommand\algorithmicassert{\texttt{assert}}
\algnewcommand\Assert[1]{\State \algorithmicassert(#1)}%
% New "environments"
\algdef{SE}[SWITCH]{Switch}{EndSwitch}[1]{\algorithmicswitch\ #1\ \algorithmicdo}{\algorithmicend\ \algorithmicswitch}%
\algdef{SE}[CASE]{Case}{EndCase}[1]{\algorithmiccase\ #1}{\algorithmicend\ \algorithmiccase}%
\algtext*{EndSwitch}%
\algtext*{EndCase}%

%New command for correct alignment for circled numbers.
% \newcommand*\circled[1]{\tikz[baseline=(char.base)]{
            % \node[shape=circle,draw,inner sep=2pt] (char) {#1};}}

\newcommand{\fref}[1]{\mbox{Figure~\ref{#1}}}
\newcommand{\lref}[1]{\mbox{Listing~\ref{#1}}}
\newcommand{\thr}[1]{\textcolor{red}{#1}}
\newcommand{\thb}[1]{\textcolor{blue}{#1}}
\newcommand{\secref}[1]{{\S\ref{#1}}}
\newcommand{\tref}[1]{\mbox{Table~\ref{#1}}}
\newcommand{\rc}[1]{{\color{red}{#1}}}
\newcommand{\bc}[1]{{\color{blue}{#1}}}
\newcommand{\lc}[1]{{\color{cyan}{#1}}}

\newcommand{\mypara}[1]{\smallskip\noindent\emph{#1}\xspace}
\newcommand{\myparab}[1]{\vspace{0.025in}\noindent\textbf{#1}}
\newcommand{\myparasc}[1]{\smallskip\noindent\textsc{#1}\xspace}
\newcommand{\note}[1]{\textbf{\color{red}{[#1]}}}
\newcommand{\eg}{{\it e.g.}\xspace}
\newcommand{\ie}{{\it i.e.}\xspace}
\newcommand{\etal}{{\em et al.}\xspace}
\newcommand{\eat}[1]{}
\newcommand{\arch}{{Yaksha}\xspace}
\newcommand{\archp}{{Yaksha-Prashna}\xspace}
\newcommand{\pt}[1]{{\color{cyan}{PT$\bigstar$:#1}}}

%Useful macros for various keywords used in the paper:
\newcommand{\Infrastructure}{Our infrastructure}
\newcommand{\infrastructure}{our infrastructure}
\newcommand{\Oursystem}{Our system}
\newcommand{\oursystem}{{our system}\xspace}
\newcommand{\system}{{System}\xspace}

\newcommand{\pluseq}{\mathbin{{+}{=}}}

\newcommand{\concateq}{\mathbin{{\parallel}{=}}}

\newcommand{\regsrc}{\mathcal{R}[src]}
\newcommand{\regdst}{\mathcal{R}\left[dst\right ]}
\newcommand{\mem}{\mathcal{M}}
\newcommand{\context}[1]{\mathcal{C}\left[#1\right]}

\newcommand{\inst}[1]{\text{I}: #1}
\newcommand{\customvalue}[1]{\text{\textbf{{#1}}}}
\newcommand{\custvariable}[1]{#1}

\newcommand{\ruleref}[1]{Rule (\ref{#1})}

\newcommand{\srcregtag}{source register \textit{tag}}
\newcommand{\srcregval}{source register \textit{value}}

\newcommand{\dstregtag}{destination register \textit{tag}}
\newcommand{\dstregval}{destination register \textit{value}}
\newcommand{\qpred}{Q-Pred\xspace}
\newcommand{\qpreds}{Q-Preds\xspace}
\newcommand{\qpredlong}{Q-Predicate\xspace}
\newcommand{\qpredslong}{Q-Predicates\xspace}
\newcommand{\rpredlong}{R-Predicate\xspace}
\newcommand{\rpred}{R-Pred\xspace}
\newcommand{\rpreds}{R-Preds\xspace}
\newcommand*{\Comb}[2]{{}^{#1}C_{#2}}%

\newcommand{\CFG}{{CFG-NC}\xspace}
\newcommand{\ControlFlowGraph}{{Control Flow Graph with Network Context}\xspace}

\newcommand{\crossmark}{%
\tikz[scale=0.23, rotate=10] {
    % line /
    \draw[line width=0.6,line cap=round] (0.2,0.2) to [bend left=10] (1.05,1.05);
    % line \
    \draw[line width=0.6,line cap=round] (0.5,1) to [bend right=3] (0.8,0.2);
}}

\newcommand{\halftick}{%
  % \tikz[scale=0.3]{\draw (0,0) -- (0.5,0.5);}} 
  \textcolor{black}{\checkmark}{\small\textcolor{black}{\kern-0.62em\crossmark}}}

\title{Yaksha-Prashna: Understanding eBPF Bytecode Network Function Behavior}
% \author{Animesh Singh}

%
% The "author" command and its associated commands are used to define
% the authors and their affiliations.
% Of note is the shared affiliation of the first two authors, and the
% "authornote" and "authornotemark" commands
% used to denote shared contribution to the research.
\author{Animesh Singh}
\affiliation{%
  \institution{Indian Institute of Technology Hyderabad}
  \country{}
  % \city{Hyderabad}
  % \postcode{502 285}
}
% \email{cs22m24p100001@iith.ac.in}

\author{K Shiv Kumar}
\affiliation{%
  \institution{Indian Institute of Technology Hyderabad}
  \country{}
}
% \email{cs21resch11003@iith.ac.in}

\author{S. VenkataKeerthy}
\affiliation{%
  \institution{Indian Institute of Technology Hyderabad}
  \country{}
}
% \email{cs17m20p100001@iith.ac.in}

\author{Pragna Mamidipaka}
\affiliation{%
  \institution{Carnegie Mellon University}
  \country{}
}
% \email{pmamidip@andrew.cmu.edu}

\author{R V B R N Aaseesh}
\affiliation{%
  \institution{Indian Institute of Technology Hyderabad}
  \country{}
}
% \email{aaseesh.rallapalli@gmail.com}

\author{Sayandeep Sen}
\affiliation{%
  \institution{IBM Research India}
  \country{}
}
% \email{sayandes@in.ibm.com}

\author{Palanivel Kodeswaran}
\affiliation{%
  \institution{Walmart Global Technology Labs}
  \country{}
}
% \email{Palani.k@gmail.com}

\author{Theophilus A. Benson}
\affiliation{%
  \institution{Carnegie Mellon University}
  \country{}
}
% \email{theophilus@cmu.edu}

\author{Ramakrishna Upadrasta}
\affiliation{%
  \institution{Indian Institute of Technology Hyderabad}
  \country{}
}
% \email{ramakrishna@cse.iith.ac.in}

\author{Praveen Tammana}
\affiliation{%
  \institution{Indian Institute of Technology Hyderabad}
  \country{}
}
% \email{praveent@cse.iith.ac.in}


\begin{abstract}
Many cloud infrastructure organizations increasingly rely on third-party eBPF-based network functions for use cases like security, observability, and load balancing, so that not everyone requires a team of highly skilled eBPF experts. However, the network functions from third parties (e.g., F5, Palo Alto) are available in bytecode format to cloud operators, giving little or no understanding of their functional correctness and interaction with other network functions in a chain. Also, eBPF developers want to provide proof of functional correctness for their developed network functions without disclosing the source code to the operators.

We design \archp, a system that allows operators/developers to assert and query bytecode's conformance to its specification and dependencies on other bytecodes. Our work builds domain-specific models that enable us to employ scalable program analysis to extract and model eBPF programs. 
Using \archp language, we express $24$ properties on standard and non-standard eBPF-based network functions with 200-1000x speedup over the state-of-the-art work.

\par

\end{abstract}


\maketitle

\section{Introduction}
\label{sec:introduction}

Today, the cloud infrastructure of many large and prominent organizations~\cite{google_ebpf, meta_ebpf, ibm_ebpf, cloudfare_ebpf} rely on eBPF-based network functions (NFs) for critical management functionality (e.g., security~\cite{hXDP}, observability~\cite{cilium}, and load balancer~\cite{katran}). These eBPF network functions have been the cause of several high profile outages, e.g., Datadog's 5 Million Dollar outage~\cite{datadog_ebpf_journey, datadog_5mdollar_outage} or outages discussed in Meta's NetEdit~\cite{netedit:sigcomm24}. While eBPF has been adopted because of its programmability, high performance, and safety guarantees, eBPF development requires a team of highly-skilled experts~\cite{ebpf_extensions_behl2023ebpf}, forcing many organizations to rely on eBPF programs from third parties (e.g., F5's DoS protect~\cite{f5_nginx_app_protect}, Palo Alto's firewall~\cite{palo_alto_cn_series_firewall}, Datadog~\cite{datadog_npm}, Cilium~\cite{cilium}, and Meta's load balancer (katran)~\cite{katran}). Due to recent large-scale outages caused by such third-party NFs, there is a growing concern about the functional correctness of these eBPF programs~\cite{draco_lu2024towards, iyer22pix,l3af}.  

Unfortunately, understanding the functional correctness and their interaction with other NFs is non-trivial because it requires understanding program behavior which varies from kernel to kernel~\cite{netedit:sigcomm24}, requires analyzing program code which includes many BPF-specific constructs and lastly documentation of the eBPF language and programs are poor. Even advanced developers, struggle because understanding NF behavior requires both interactions between the NF and other NFs deployed along side but also with the kernel. 
More importantly, to check an NF's conformance with its specification, it is crucial to know the network context, such as read/write/copy operations on packet content, protocols processed, and eBPF maps invoked.  The bytecode's network context also helps to understand its interaction with other NFs (\eg, Write before Read on a packet header). 
Thus, while understanding eBPF-based functions gives operators confidence to debug or avoid outages before deployment in production networks~\cite{cilium_identity_overwrite, ebpf_murder_mystery}, they lack such knowledge and may have to wait for a longer duration~\cite{empirical_study_deokar2024empirical} for support from third-party eBPF program developers.

In this paper, we take a classic diagnosis approach to the problem: we aim to hide the nuances associated with kernel versions, eBPF specification, detailed knowledge of program design and interactions by introducing a query language that \textit{enables network operators and developers to assert and query eBPF-based NFs from their bytecode and understand their behavior}.
Our query language enables a broad set of analysis through simple language constructs (e.g., pre/post-conditions~\cite{gries1981science})
and language extensions (e.g., writing assertions/contracts~\cite{stroustrup2022tour}).
To support this query language we introduce a program analysis-based (e.g., program verification~\cite{hoare69,floyd1993verification} and abstract interpretation~\cite{cousot-cousot-1977-abstractinterpretation,cousot-halbwachs-1978}) framework that builds models that infer a program's behavior. 
Our work differs from prior work on diagnosis because of our need to address eBPF specific constructs, e.g., helpers and maps, which make eBPF bytecode sufficiently different from traditional programs. To this end, we need to address two novel challenges: an eBPF-centric static analysis model, and an eBPF-centric query system to support user interaction.

\myparab{Behavioral model for analysis.} Existing works on NF verification rely on formal models of the source code for verifying the properties of the NFs implemented in high-level languages (like C and P4). 
For instance, tools like eBPF-SE~\cite{iyer22pix} model eBPF programs and leverage the generic symbolic execution engines like Klee~\cite{cadar08klee} to analyze these models. 
However, the applicability of such works to eBPF bytecode is limited due to a lack of behavioral models on eBPF bytecode. On the other hand, there are eBPF verifiers~\cite{verifier, prevail} that operate on bytecode, thus alleviating the need for source code, but they mainly model the semantic properties to validate program termination, memory safety, and resource boundedness, using static analysis~\cite{verifier}, abstract interpretation~\cite{prevail} and symbolic execution~\cite{verify_nf_binaries} techniques. These verifiers are not designed to extract network context from bytecode.

Extracting network context from a bytecode (\eg, protocol processed) introduces two key challenges over the standard source-code analysis: First, eBPF bytecode instructions are at a low level, similar to assembly. Each high-level statement can translate into multiple bytecode instructions, making it challenging to interpret and understand the relation to other instructions in the execution flow. Second, due to the limited number of registers and stack memory, large and complex eBPF programs reuse registers and memory locations, leading to frequent redefinitions that further obscure the analysis.

Addressing these challenges, we design \textbf{\archp Analyzer}, which analyzes bytecode based on the classic dataflow analysis~\cite{hecht-book-77}, guided by control flow analysis (\secref{subsec:cfg_nc}). 
The analyzer captures granular information at each program point by carefully tracking data flow across registers and memory locations (\secref{subsec:data_structure}). We define dataflow rules to extract the network context of the eBPF bytecode effectively (\secref{subsec:nce}). 
Interestingly, the characteristic features of eBPF, like limited state space, absence of loops, and a fixed number of registers, etc., make our analyzer both efficient and scalable.

\myparab{Behavioral domain-specific language (DSL).} Another requirement is the user interaction by expressing behavioral queries on the eBPF bytecode. 
Existing tools for NF verification take the standard program verification approach and primarily focus on extending the (source) language with assertions~\cite{draco_lu2024towards,p4VerifcnAssert,p4assert,sdn_assert}. These extended assertion languages have limited scope; Assertions are good at validating if certain properties hold (e.g., ``\textit{Does the NF write to a map?}'') but cannot retrieve the network context (e.g., ``\textit{Which packet header field is written to the map by NF?}''). Furthermore, each assertion typically requires its separate analysis pass; therefore, for each new assertion, the program is analyzed from scratch to validate those additional properties, thereby incurring significant computational overheads (both in analysis time and memory consumption). 

Consequently, there is a necessity for a custom query language that decouples the retrieval of network context from querying the extracted network context. Designing this query language poses two main challenges: First, the language should effectively be able to map the high-level intention of NF operators/developers to the low-level constructs in the bytecode while effectively abstracting out the low-level details in the bytecode. Second, the language should be able to express complex queries that check conformance and also bytecode's interaction with other NFs.

Addressing these challenges, we design \textbf{\archp Language}, a simple but expressive DSL, to abstract the complexities of the analysis. 
\archp allows network operators/developers to express high-level queries about NF behavior without lower-level bytecode details like the basic block IDs, register mappings, etc. 
\archp supports users to \textit{express} two types of queries, assertion and retrieval, on individual bytecode and across multiple bytecodes for cross-function interactions. It is \textit{composable}, allowing users to combine simple queries to formulate complex real-world queries, and \textit{adoptable}, leveraging Prolog for easy logic-based querying.

To ensure \textit{scalability} of our system, we decouple the bytecode analysis from the query phase, allowing the bytecode to be analyzed once, with the extracted network context reused to answer any subsequent queries. This approach eliminates repeated analysis passes, reducing overhead and supporting efficient, large-scale deployment.

\noindent The key contributions of our paper are as follows:
% \vspace{-1mm}
\begin{itemize}[leftmargin=*]
    \item We propose \archp Language, a domain-specific query language for NF operators/NF developers to assert and retrieve network context of eBPF bytecode, abstracting the complex dependencies among instructions in low-level bytecode (\secref{sec:query_engine}).

    \item We construct a formal dataflow analysis model, guided by control flow analysis, for eBPF bytecodes and design \archp Analyzer to extract network context (\secref{sec:bca}).

    \item We present \archp (\secref{sec:overview}), a comprehensive system that integrates the analyzer (\secref{sec:bca}) and the query language (\secref{sec:query_engine}). The system is scalable as it decouples the bytecode analysis from the query execution phase.

    \item We prototype \archp and tested by executing queries on $16$ XDP programs from projects like Katran~\cite{katran} and Suricata~\cite{suricata}. 

    The analyzer extracts network context of bytecode with $4$K paths in $300$ $ms$ while consuming less than $50$ MB of memory. \archp's query system executes $15$ queries on $16$ XDP programs in less than $50$ $\mu$s while consuming less than $15$ MB of memory. 

    \item We compared \archp with state-of-the-art tools, Klint~\cite{verify_nf_binaries} and DRACO~\cite{draco_lu2024towards}, based on expressivity of the language and verification time. \archp can express the properties used for verification while keeping the verification time $200$-$1000$$\times$ less.

\end{itemize}

\section{Background and Motivation}\label{sec:background}

\subsection{Third-party eBPF network functions}

Many organizations rely on third-party eBPF programs.  While many are open-source (\eg, Data-dog's NPM, Cilium's CNI, Meta's Katran), a growing number of observability and security programs are proprietary, and their bytecodes are distributed to operators either directly~\cite{f5_nginx_using_eBPF, palo_alto_cn_series_firewall} or through marketplaces like L3AF~\cite{l3af}. Some example closed and proprietary eBPF-based NFs are NGINX App protect DoS from F5 Networks~\cite{f5_nginx_app_protect} and CN-series firewall by Palo Alto~\cite{palo_alto_cn_series_firewall}. We see this is in line with the marketplace paradigm for virtual network functions (VNF) like equinox network edge~\cite{equinix_network_edge} which offers VNFs from various vendors like Cisco, Juniper, Palo Alto, and Fortinet. 

\myparab{Why understanding bytecode is important?} Typically, an eBPF-based NF is developed in a high-level language~\cite{c_language_kernighan2002c, cpp_language_stroustrup1994design, python_language_sanner1999, rust_language_matsakis2014rust, lua_langauge_ierusalimschy2006programming, go_language_donovan2015go} and is compiled into bytecode using Clang/LLVM toolchain~\cite{llvm_compiler_infra}. 
The bytecode is loaded into the host kernel with the help of system calls~\cite{bpf_syscall}. During this process, the eBPF verifier performs static checks on the bytecode to ensure that the bytecode runs safely inside the kernel. 
However, the \textit{verifier does not model or reason the functional correctness of the bytecode}. More specifically, the third-party eBPF-based NF may not be buggy, but it may not fully cover the target behavior~\cite{draco_lu2024towards}, that is, functional correctness. This can happen due to various reasons, such as missing statements (e.g., a missing TTL decrement in a router \cite{aquila}), unexpected packet forwarding behavior (e.g., router forwarding packets with TTL = 0), or deviations from standard NF behavior (e.g., a firewall unintentionally allowing unknown traffic), etc. Also, the NF's interactions with other NFs in the system can cause unexpected behavior. If the programs are from eBPF marketplaces~\cite{l3af}, they may have unknown or malicious code~\cite{draco_lu2024towards}.  This motivates the need to understand the bytecode's packet-processing behavior and its interactions with other NFs before deploying it in the network. 

\myparab{Intented users of \archp.} Network operators and network developers are the two main users of the proposed tool. \textbf{(1) Network operators:} The documentation and specification for closed bytecodes give little visibility into the inner workings~\cite{verify_nf_binaries}. An NF's specification informs what the NF can do, but does not specify how the NF should be implemented. If the NF conforms to the respective specification, then the operator is confident to deploy it, even though access to the source code is restricted. Moreover, operators may want to verify that integrating new NF will not cause disruptions or unintended interactions within the existing NF chains. Even when source code is provided, as with Cilium, understanding such behaviors requires tedious and significant efforts. \archp enables operators to validate the functional correctness of NFs to be deployed safely in the existing NF chain.
\textbf{(2) Network developers:} The developers want to guarantee the functional correctness of their developed NFs, without disclosing the source code, to the operators.
Moreover, bytecode conformance to specification enables developers to use any language and toolchain, especially those in the experimental phase, so that the compiled binary satisfies the target behavior~\cite{verify_nf_binaries}. \archp helps the developers in achieving these goals.

%############################ Version 2 ############################

\begin{figure}[t]
    % \centering
    % \includegraphics{}
\begin{framed} 
% \begin{center}
\resizebox{1.1\columnwidth}{!}{%
\scriptsize
\begin{bnf}[
colspec = {lrcll}, 
column{1} = {font = \sffamily},
column{2} = {mode = dmath},
column{4} = {font = \sffamily},
column{5} = {font = \it\color{green! 45! black}}
]

% This grammar require simplebnf.sty to be included in the over leaf repo.

Query ::= \texttt{Q-Predicate\_list} \texttt{`.'} ;;

Q-Predicate\_list ::= \texttt{Q-Predicate} \texttt{[}\texttt{operator} \texttt{Q-Predicate]}\texttt{*}   ;;

% \bc{Quantifier} ::= \bc{\texttt{forall}};;
Q-Predicate ::=	\texttt{`!'}\texttt{`('}\texttt{Q-Predicate}\texttt{`)'} | \texttt{field\_pred}\texttt{`('} \texttt{nf\_id} \texttt{`,'} \texttt{fld\_arg}\texttt{`)'} 
				   | \texttt{pkt\_act\_pred}\texttt{`('} \texttt{nf\_id} \texttt{`,'} \texttt{hook\_arg}  \texttt{`,'} \texttt{list\_of\_pairs}\texttt{`)'} 
                      % | \textit{\rc{map\_pred}}\texttt{`('} \textit{nf\_id} \texttt{`,'} \textit{map\_id} \texttt{`,'} \textit{fld\_arg}\texttt{`)'}
                      % | \textit{map\_lookup}\texttt{`('} \textit{nf\_id} \texttt{`,'} \textit{map\_id} \texttt{`)'}
                      | \texttt{map\_operation\_pred}\texttt{`('}\texttt{nf\_id}\texttt{`,'}\texttt{map\_id}\texttt{`,'}\texttt{fld\_arg}\texttt{`)'}
                      | \texttt{correlated\_maps\_pred}\texttt{`('}\texttt{nf\_id}\texttt{`,'}\texttt{list\_of\_maps}\texttt{`)'}
                      | \texttt{order\_pred}\texttt{`('} \texttt{nf\_id} \texttt{`,'} \texttt{var}\texttt{`)'} 
                      | \texttt{protocol\_pred}\texttt{`('} \texttt{nf\_id} \texttt{`,'}\texttt{fld\_arg} \texttt{`,'}\texttt{value}\texttt{`)'} 
                      | \texttt{helper\_pred}\texttt{`('} \texttt{nf\_id} \texttt{`,'} \texttt{helpers}\texttt{`)'} 
                      ;;

% function  ::=  \textit{f\_order\_nf} \textbf{`('} \textit{nf\_arg} \textbf{`,'} \textit{var} \textbf{`)'} ;;
operator ::= \texttt{`,'} // \texttt{`;'} ;;
% list\_of\_pairs ::= \texttt{[\texttt{`('}\textit{fld\_arg} \texttt{`,'} \textit{value} \texttt{`)'}]}  ;;

list\_of\_pairs ::= \texttt{`['}\texttt{pair} \texttt{[}\texttt{operator} \texttt{pair}\texttt{]*}\texttt{`]'} ;;

list\_of\_maps ::= \texttt{\texttt{`['}\texttt{`('}\texttt{map\_id} \texttt{`,'} \texttt{map\_id} \texttt{`)'}\texttt{`]'}}  ;;

pair ::= \texttt{`('}\texttt{fld\_arg} \texttt{`,'} \texttt{value} \texttt{`)'}  ;;

field\_pred  ::= \texttt{updatesField} // \texttt{readsField} ;;

pkt\_act\_pred ::= \texttt{drops} // \texttt{passes} // \texttt{aborts} // \texttt{redirects} // \texttt{tx} // \texttt{all} ;;

map\_operation\_pred ::= \texttt{mapLookup} // \texttt{mapWrite} ;;

% map\_write ::= \texttt{writesToMap} ;;

correlated\_maps\_pred ::= \texttt{correlatedMaps} ;;

order\_pred ::=	\texttt{successorNF} //  \texttt{predecessorNF} ;;

protocol\_pred  ::= \texttt{accessesProtocol} ;;

helper\_pred  ::= \texttt{callsHelper} ;;

% protocol ::=  \texttt{eth} // \texttt{ipv4} // \texttt{ipv6} // \texttt{tcp} // \texttt{\dots};;

fld\_arg ::= \texttt{header\_field} // \texttt{buffer\_field} // \texttt{var} // \texttt{`*'};;

hook\_arg ::=  \texttt{xdp} // \texttt{tc} // \texttt{var} // \texttt{\dots} : \hspace{-2.5cm} /\ /\ Hook points ;;

header\_field ::= 	\texttt{eth.type} // \texttt{eth.dst} // \texttt{eth.src} // \texttt{\dots} | \texttt{ipv4.src} // \texttt{ipv4.dst} // \texttt{\dots}  | \texttt{\dots} : \hspace{-3cm} /\ /\ Std header fields ;;

buffer\_field ::=	\texttt{xdp\_md.data} // \texttt{xdp\_md.data\_end} // \texttt{\dots} | \texttt{sk\_buff.mark} // \texttt{\dots} | \texttt{\dots} : \hspace{-3.3cm} /\ /\ Kernel Buffer fields;;

helpers ::= \texttt{bpf\_map\_lookup\_elem} | \texttt{bpf\_map\_update\_elem} | \texttt{bpf\_probe\_write\_user}// \texttt{var} // \texttt{\dots} | \texttt{\dots} : \hspace{-3.5cm}
/\ /\ eBPF helper functions;;

nf\_id ::= \texttt{string} // \texttt{var} // \texttt{`*'} ;;

map\_id ::= \texttt{string} // \texttt{var} // \texttt{`*'} ;; 

value ::=  \texttt{const} // \texttt{var} // \texttt{`*'}  ;; 

var ::= \texttt{[A-Z\_][A-Za-z0-9\_]}* : \hspace{-3.5cm}
/\ /\ Variable ;;

const ::= \texttt{!const} // \texttt{[0-9]}\texttt{+} // \texttt{string} // \texttt{ip-address}
\end{bnf}
% \end{center}
 }
\end{framed}

    \caption{Grammar for \archp Language.}
    % \urkfixme{HERE AND ELSEWHERE WE SHOULD use \arch or eBCQL }
    % \sk{Remove field names, align it horizontally.}
    \label{fig:grammar}
    % \Description{Grammar for Yaksha}
    \vspace{-0.3cm}
\end{figure}

%############################  GRAMMAR 
%\eat{\section{Example assertions using \arch-Language}
%\label{sec:grammar}

%closed or open

%%%%%%%%%%%%%%%%%%%%%%%%%%%%%%%%%%%%%%%%%%%%%%%%%%%%%%%%%%%%%%%%%%%%%%%%%%%%%%%%%%%%%%%%%%%%%%%%%%%%
\subsection{Motivating use cases}
\label{subsec:motivating_examples}

A usable verification system should meet three requirements \cite{aquila} for users so that they can: (1) express the correct NF behavior (specification) with ease, (2) check the conformance of the NF/NF chain within a reasonable time and low memory overhead, and (3) reason about the underlying causes of specification violations. \archp language (\fref{fig:grammar}) is designed to meet all three, so that NF operators/NF developers can validate bytecode using assertions and understand the NF behavior using retrieval queries. 
Details of the language are in Section~\secref{sec:query_engine}.

We categorize the use cases of \archp into three.

\myparab{Category 1: Individual NF conformance to specification.}  Consider that a user (NF operator/NF developer) wants to validate whether an NF implementation (bytecode) conforms to its specification. For example, a stateful firewall specification says (1) it should not modify packet contents; and (2) it should allow only TCP and UDP traffic, and drop other traffic (\eg, ICMP~\cite{draco_lu2024towards}). Listing \ref{lst:firewall_snippet} shows an example stateful firewall code snippet that does not conform to the specification; it updates the source port (line 10) of TCP traffic and processes ICMP traffic (line 6). They must know that the bytecode generated from this eBPF program does not conform to the specification. To do so, they can write the following assertions to check whether the NF bytecode conforms to the specification.

The assertion below asserts that the firewall bytecode (Listing~\ref{lst:firewall_snippet}) does not modify (update) packet content.
      \begin{center}
      \textbf{A1: }\texttt{!updatesField(xdp\_fw, *).}
      \end{center}
\noindent It gets the list of packet fields updated by the bytecode and returns false if the list is not empty. The argument $*$ indicates that this is an assertion. The assertion fails on the firewall because it modifies the TCP source port.

The assertion below checks that the firewall bytecode at the XDP hook point does not allow the ICMP traffic. 
      \begin{center}
      \textbf{A2: }\texttt{passes(xdp\_fw, xdp, [(var, var)]), !accessesProtocol(xdp\_fw, ``ipv4.proto'', ``icmp'').}
      \end{center}
    
\noindent $passes$ construct with ($var$, $var$) as an argument returns the network context of all program paths that passes the packet. Next, it checks that these paths should not access the ICMP protocol headers, using $!accessesProtocol$ construct. The assertion fails on the bytecode generated from Listing~\ref{lst:firewall_snippet} because it processes ICMP traffic (line 6), and passes the packet.

%%%%%%%%%%%% CODE Snippets %%%%%%%%%%%%%%%%%%%%%%
\begin{lstlisting}[belowskip=-3.5mm, float=tp, caption={Firewall code snippet}, label={lst:firewall_snippet}]
int xdp_fw(struct xdp_md *ctx) {
...
switch(ip->protocol){
  case IPPROTO_TCP :  goto l4;
  case IPPROTO_UDP :  goto l4;
  @case IPPROTO_ICMP : goto l4;@ /* Parses ICMP */
  default :           goto EOP;
}
...
@tcp->sourceport = 8080;@ /* updates TCP source port */
...
if(ingress_ifindex == EXTERNAL){
  flow_leaf = bpf_map_lookup_elem(&flow_ctx_table, &flow_key);
  if(flow_leaf) 
    return bpf_redirect_map(&tx_port, flow_leaf->out_port, 0);
... }
else{
flow_leaf = bpf_map_lookup_elem(&flow_ctx_table, &flow_key);
		if (!flow_leaf){ ...
			bpf_map_update_elem(&flow_ctx_table, &flow_key, &new_flow, 0);}
...}
EOP : return XDP_DROP;
}
\
\end{lstlisting}

%%%%%%%%%%%%%%%%%%%%%%%%%%%%%%%%%%
%###################### Listing 2 starts #########################
\begin{lstlisting}[belowskip=-3.5mm, float=tp, label={lst:update_mark}, caption={Unintended write by AWS's NF2 before read by Cilium's NF3 lead to traffic drop.}]
    ?*********** Cilium NF (NF1) ***********?
int identity_manager(struct __sk_buff *skb) {
  __u32 *secid = bpf_map_lookup_elem(&sec_id_map, &key);
...
  /* Extract first 8 bits (most significant byte) */
  __u8 id_prefix = (*secid >> 24) & 0xFF; 
  if(skb->mark){
    /* Updates skb->mark field */
    @skb->mark = (skb->mark & 0xFFFFFF00) | id_prefix;@
  }
... }

    @*********** AWS NF (NF2) ***********@
int pbr_routing(struct __sk_buff *skb) {
  ...
  *masks = bpf_map_lookup_elem(&mask_map, &key);
  ...  
  __u32 new_mark = (nfmark & ~nfmask) ^ (CTMARK & ctmask);
  @skb->mark = new_mark;@ /*Updates skb->mark field*/
  ... }

    ?*********** Cilium NF (NF3) ***********?
int enforce_policy(struct __sk_buff *skb) {
	...
	/* Retrieves Cluster_id from skb->mark field */
    @__u32 cluster_id = skb->mark & 0xFF;@
    __u32 *val = bpf_map_lookup_elem(&cluster_map, &cluster_id);
	/* Cluster ID not found, drop the packet */
    if (!val) { return TC_ACT_SHOT;}
    /* Cluster ID is known, allow packet */
    return TC_ACT_OK;
}
\end{lstlisting}
%###################### Listing 2 ends #########################

\myparab{Category 2: Unintended NF interactions.} Multiple eBPF programs are attached to the same hook point (\eg, XDP, TC). This enables the development of modular programs to build complex systems. However, an update to a state by one program can lead to unexpected behavior if another program in the chain relies on that state. More specifically, some dependencies that an operator may want to know are Read after Write (RAW), Write after Read (WAR), and Write after Write (WAW)~\cite{draco_lu2024towards}. To illustrate this further, consider an issue observed in practice that resembles RAW dependency. A bad interaction between Cilium and AWS elastic network interface~\cite{cilium_identity_overwrite, ebpf_murder_mystery} resulted in an unintended drop in traffic for some customers. This interaction is illustrated using a code snippet in Listing \ref{lst:update_mark}. Here, Cilium's $identity\_manager$ (NF1), stores security ID $id\_prefix$ in \textit{sk\_buff->mark} field to represent the traffic from a specific pod in a specific cluster. Down the chain, another NF from cilium, $enforce\_policy$ (NF3), does a lookup on $cluster\_id$ in \textit{sk\_buff->mark} (line 26-32) and allows traffic. However, $pbr\_routing$ (NF2) at the AWS elastic network interface, introduced newly into the packet path, updates \textit{sk\_buff->mark} (line 19) to capture the connection type (default). As a consequence, $cluster\_id$ is corrupted, and traffic for some customers is dropped. To avoid such bad interactions before they happen, it is crucial to know dependencies before a bytecode is introduced into a chain.

The following assertion checks for Read After Write (RAW) dependency between NF2 and any successor NFs (NF3) in the chain, as shown in Listing~\ref{lst:update_mark}. This assertion can be executed on the new chain before deploying NF2's bytecode. 
    \begin{center}
    \textbf{A3: }\texttt{!updatesField(NF2, *), successorNF(NF2, SNf), readsField(SNf, *).}
    \end{center}
$updatesField$ with $*$ argument retrieves a list of fields updated by NF2's bytecode (write list), including socket buffer and packet headers. $successorNF$ gets all the successor NFs of NF2, and $readsField$ construct retrieves the list of fields read by them (read list). The assertion fails if the write list overlaps with the read list of any of NF2's successors in \texttt{SNf}.
\archp language also supports assertions to check for other dependencies (WAR, WAW).

\myparab{Category 3: Understanding the NF/NF chain behavior.} Once it has been established that the NF bytecode does not conform to its specification, or that a dependency exists in an NF chain, identifying the cause enables precise and reliable bug reporting. For example, one can accurately document and report that the specification violation in Listing~\ref{lst:firewall_snippet} was due to the modification of the TCP source port, and \textit{sk\_buff->mark} field is responsible for the dependency in Listing \ref{lst:update_mark}.

In the first case, the retrieval query 
      \begin{center}
      \textbf{RQ1: }\texttt{updatesField(xdp\_fw, Fld).}
      \end{center}
%\noindent It 
retrieves a list of packet fields (\eg, \textit{tcp.sport}) updated by the firewall, stores them in $Fld$, and returns as the query result.

In the second case, the following retrieval query can be used to debug the assertion A2's failure.
    \begin{center}
    \textbf{RQ2: }\texttt{updatesField(NF2, Fld), successorNF(NF2, SNf), readsField(SNf, Fld).}
    \end{center}
\noindent It retrieves the list of socket buffer and packet header fields updated by NF2 and read by any of NF2's successors, stores them in $Fld$, and returns them as the query result (\ie, \textit{sk\_buff->mark} in Listing \ref{lst:update_mark}).
To summarize, retrieval queries provide deeper insight into packet processing behavior, such as protocols being accessed, header/buffer fields being read/updated, shared map between NFs, helper functions invoked, etc.

\section{\archp overview}\label{sec:overview}

\begin{figure}[t]
     \centering
     \includegraphics[width=0.48\textwidth]{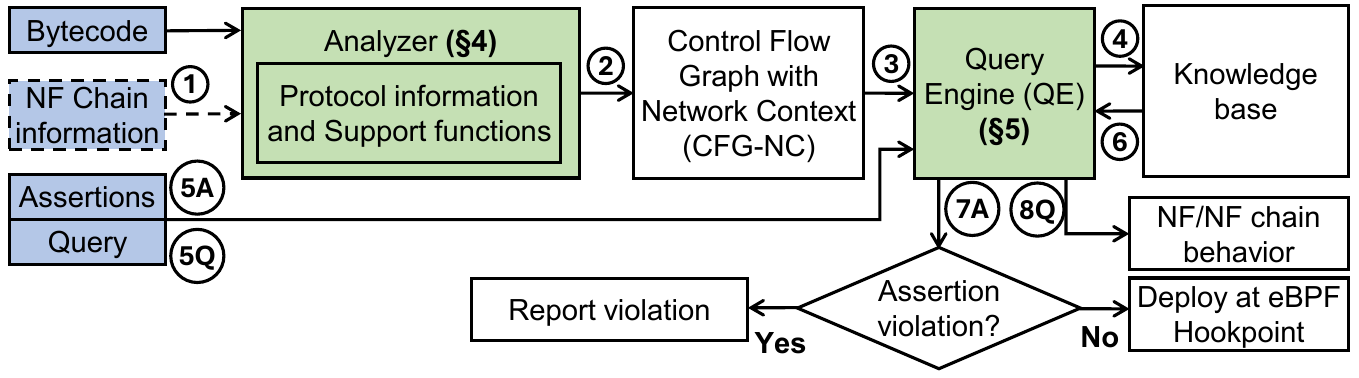}
     \caption{\archp system workflow}
     \label{system_flow}
     % \Description{\arch Implementation}
     \vspace{-6mm}
 \end{figure}

\archp workflow is presented in~\fref{system_flow}. It consists of two main components: the \textbf{Analyzer} (\secref{sec:bca}) and the \textbf{Query Engine (QE)} (\secref{sec:query_engine}). \textbf{(1)} For a given bytecode (optionally accompanied by NF chain information, if the user wants to check for unintended interactions before deploying a new NF), the analyzer builds a formal dataflow analysis model guided by control flow analysis. This model is represented as a control flow graph (CFG) and annotated with the network context (NC) extracted from the bytecode, resulting in \textbf{(2)} the \CFG. We define the bytecode’s network context as the set of read/write/copy operations on packet content (headers and metadata), the network protocols processed, the eBPF maps accessed, and the helper functions invoked by the bytecode. \textbf{(3)} The Query Engine (QE) translates the network context encoded in the CFG-NC into \textbf{(4)} the Knowledge Base (Prolog facts), while preserving the structural integrity of the \CFG. \textbf{(5A)} Assertions written in the language (Figure \ref{fig:grammar}) are evaluated by the QE by \textbf{(6)} mapping their predicates to the corresponding facts stored in the Knowledge Base. \textbf{(7A)} If an assertion violation is detected, it is reported; otherwise, the NF can be deployed at the eBPF hook point. In the case of a violation, \textbf{(5Q)} retrieval queries are issued to the QE, and \textbf{(8Q)} the behavior of the NF or NF chain is retrieved.

% ####################################### transfer function rules START ##################################
\begin{table*}[t]
% \footnotesize
\scriptsize
\centering
\caption{Transfer function rules.}
\label{tab:inference_rules1}
\resizebox{1\linewidth}{!}{%
\centering
% \begin{tabular}{lclclclcl}
\begin{tabular}{lllll}
\toprule
% \cmidrule{5-6}
% \multicolumn{1}{c}{\multirow{2}{*}{\textbf{\begin{tabular}[c]{@{}l@{}}Rule\\ No.\end{tabular}}}} &
\multicolumn{1}{c}{\multirow{2}{*}{\textbf{\begin{tabular}[c]{@{}l@{}}Rule\end{tabular}}}} 
& \multicolumn{1}{c}{\multirow{2}{*}{\textbf{\begin{tabular}[c]{@{}l@{}}Instruction\end{tabular}}}} 
& \multicolumn{1}{c}{\multirow{2}{*}{\textbf{\begin{tabular}[c]{@{}l@{}}Condition\end{tabular}}}} 
& \multicolumn{2}{c}{\textbf{State transformation} } \\ \cline{4-5}

% \multicolumn{1}{c}{} &
\multicolumn{1}{c}{}                                 
& \multicolumn{1}{c}{}                                                                                   
& \multicolumn{1}{c}{}                                        
& \multicolumn{1}{c}{\textbf{Reg. and Mem. states}} 
& \multicolumn{1}{c}{\textbf{Context state}} \\ \toprule

% Rules
\begin{tabular}[c]{@{}l@{}} 
\textbf{R1:} Arithmetic \\ operation
\end{tabular}
& \begin{tabular}[c]{@{}l@{}}
\( \displaystyle
% dst \leftarrow dst<\!\!op\!\!>
dst \leftarrow dst \oplus \custvariable{imm} |\ src\
\)\\
% \( \displaystyle
 
% \)
\end{tabular}

& \multicolumn{1}{c}{\textbf{--}} 
% \begin{tabular}[c]{@{}c@{}}
% \multicolumn{1}{c}{\textbf{--}} 
% \end{tabular}

& \begin{tabular}[c]{@{}l@{}}
\( \displaystyle
% \regdst.val \leftarrow  \regdst.val<\!\!op\!\!>
\regdst.val \leftarrow  \regdst.val \oplus \custvariable{imm} 
\)\\
\( \displaystyle
|\ \regsrc.val\
\)
\end{tabular}

& \multicolumn{1}{c}{\textbf{--}} 
% \begin{tabular}[c]{@{}l@{}} 
% \multicolumn{1}{l}{\textbf{--}} 
% \end{tabular}
\\ \midrule

\begin{tabular}[c]{@{}l@{}} 
\textbf{R2:} Assignment \\operation 
\end{tabular}
& \( \displaystyle
dst \leftarrow src\ |\ \custvariable{imm}
\)
& \multicolumn{1}{c}{\textbf{--}}  
% \begin{tabular}[c]{@{}l@{}}
% \multicolumn{1}{l}{\textbf{--}} 
% \end{tabular}

& \begin{tabular}[c]{@{}l@{}}
\( \displaystyle
\regdst.tag \leftarrow \regsrc.tag\ |\ \customvalue{Const}, 
\)\\
\( \displaystyle
\regdst.val \leftarrow \regsrc.val\ |\ \custvariable{imm}
\)
\end{tabular}
& \multicolumn{1}{c}{\textbf{--}} 
% \begin{tabular}[c]{@{}l@{}}
% \multicolumn{1}{l}{\textbf{--}} 
% \end{tabular}
\\ \midrule

\begin{tabular}[c]{@{}l@{}} 
 \textbf{R3:} Writing\\ to packet\\ buffer \\ \\ 

 \textbf{R4:} Writing\\ to packet data  \\ \\ 

 \textbf{R5:} Writing \\ to program \\ stack
\end{tabular}

& \begin{tabular}[c]{@{}l@{}}
\( \displaystyle
*(dst + off) \leftarrow imm\ |\ src
\)\\ \\ \\ \\ 
\( \displaystyle
*(dst + off) \leftarrow imm\ |\ src
\) \\ \\ \\ \\
\( \displaystyle
*(dst + off) \leftarrow imm\ |\ src
\) \\ 
\end{tabular}

%  ############### STORE conditions
& \begin{tabular}[c]{@{}l@{}}
\( \displaystyle
\regdst.tag = = \customvalue{pkt\_buff}
\)\\ \\ \\ \\ 
\( \displaystyle
\regdst.tag == \customvalue{pkt\_data\_start}
\)\\ \\ \\ \\
\( \displaystyle
\regdst.tag == \customvalue{stack\_frame}
\)\\
\end{tabular}

% ############# STORE REG and Mem states
& \begin{tabular}[c]{@{}l@{}}  
\multicolumn{1}{c}{\textbf{--}}  \\ \\ \\ \\  
\multicolumn{1}{c}{\textbf{--}}  \\ \\ \\ 
\( \displaystyle
\mem[off].tag \leftarrow \customvalue{Const}\ |\ \regsrc.tag,
\)\\
\( \displaystyle
\mem[off].val \leftarrow \custvariable{imm}\ |\ \regsrc.val
\)\\
\end{tabular}

% ############# STORE Context states
& \begin{tabular}[c]{@{}l@{}}
% pkt_buff
% \( \displaystyle
% \context{write\_buff}.tag \pluseq \customvalue{buff\_field},
% \)

\( \displaystyle
\context{write\_buff}.val \pluseq \big[ \text{getBuffFldName}_{spec}(
\)\\
\( \displaystyle
buff\_type,\ off), \custvariable{imm}\ |\ \regsrc.val \big]
\)\\
% \( \displaystyle

% \)
\\ \\ 

% pkt_start
% \( \displaystyle
% \context{write\_hdr}.tag \pluseq \customvalue{hdr\_field},
% \)
\( \displaystyle
\context{write\_hdr}.val \pluseq \big[\text{getHdrFldName}_{spec}(
\)\\
\(
\displaystyle
\context{curr\_proto}.val,\ off), \custvariable{imm}\ |\ \regsrc.val \big]
\)\\ 
\\ \\ 

% stack_frame
\multicolumn{1}{c}{\textbf{--}} 
\end{tabular}
\\ \midrule

% Map rule ###############
\eat{\begin{tabular}[c]{@{}l@{}} 
\textbf{R6:} Accessing \\map \\ \\ \\

\textbf{R7:} Writing \\into map
\end{tabular}

% #### Map rules: Instru
& \begin{tabular}[c]{@{}l@{}} 
\( \displaystyle
dst \leftarrow map\_by\_fd(imm64)
\) \\ \\ \\ \\  \\
\( \displaystyle
CALL\ 2
\)
\end{tabular}

% #### Map rules: Condition
% & \multicolumn{1}{c}{\textbf{--}} 
& \begin{tabular}[c]{@{}l@{}}
\multicolumn{1}{c}{\textbf{--}} \\ \\ \\ \\ \\
\( \displaystyle
\mathcal{R}[1].tag == \customvalue{$\mathbf{Ref_{map}}$}
\)
\end{tabular}
% #### Map rules: Reg and Mem states
& \begin{tabular}[c]{@{}l@{}} 
\( \displaystyle
\regdst.tag \leftarrow \customvalue{$\mathbf{Ref_{map}}$}
\)\\
\( \displaystyle
\regdst.val \leftarrow \custvariable{imm64}
\) \\ \\ \\ \\
\multicolumn{1}{c}{\textbf{--}} 
\end{tabular}

& \begin{tabular}[c]{@{}l@{}} 
% \(\displaystyle
% \context{map\_read}.tag \pluseq \mathbf{map\_read}
% \)

\(\displaystyle
\context{map\_read}.val \pluseq \text{getMapName}_{elf}()
\) \\ \\ \\

% \(\displaystyle
% \context{map\_write}.tag \pluseq \mathbf{map\_write}
% \)

\(\displaystyle
\context{map\_write}.val \pluseq [\text{getMapName}_{elf}(),
\) \\
\(\displaystyle
\text{getHdrFldName}_{spec}(\context{curr\_proto}.val,
\) \\
\(\displaystyle
\ \mathcal{R}[3].val)]
\) 
\end{tabular} 
\\ \midrule}

% Map rule ###############
\begin{tabular}[c]{@{}l@{}} 
\textbf{R6:} Accessing \\map \\ \\ \\

\textbf{R7:} Writing \\into map
\end{tabular}

% #### Map rules: Instru
& \begin{tabular}[c]{@{}l@{}} 
\( \displaystyle
dst \leftarrow map\_by\_fd(imm64)
\) \\ \\ \\ \\  \\
\( \displaystyle
CALL\ 2
\) 
\end{tabular}

% #### Map rules: Condition
% & \multicolumn{1}{c}{\textbf{--}} 
& \begin{tabular}[c]{@{}l@{}}
\multicolumn{1}{c}{\textbf{--}} \\ \\ \\ \\ \\
\( \displaystyle
\mathcal{R}[1].tag == \customvalue{$\mathbf{Ref_{map}}$}
\) 

\end{tabular}
% #### Map rules: Reg and Mem states
& \begin{tabular}[c]{@{}l@{}} 
\( \displaystyle
\regdst.tag \leftarrow \customvalue{$\mathbf{Ref_{map}}$}
\)\\
\( \displaystyle
\regdst.val \leftarrow \custvariable{imm64}
\) \\ \\ \\ \\
\multicolumn{1}{c}{\textbf{--}} 
\end{tabular}

& \begin{tabular}[c]{@{}l@{}} 
% \(\displaystyle
% \context{map\_read}.tag \pluseq \mathbf{map\_read}
% \)

\(\displaystyle
\context{map\_read}.val \pluseq \text{getMapName}_{elf}()
\) \\ \\ \\

% \(\displaystyle
% \context{map\_write}.tag \pluseq \mathbf{map\_write}
% \)

\(\displaystyle
\context{map\_write}.val \pluseq [\text{getMapName}_{elf}(),
\) \\
\(\displaystyle
\text{getHdrFldName}_{spec}(\context{curr\_proto}.val,
\) \\
\(\displaystyle
\ \mathcal{R}[3].val)]
\)
\end{tabular} 
\\ \midrule

\begin{tabular}[c]{@{}l@{}} 
\textbf{R8:} Reading \\ packet buffer \\ \\ \\ \\
\textbf{R9:} Reading \\ packet data \\ \\ \\ 
\textbf{R10:} Reading \\ program stack \\
\end{tabular}

% ################### LOAD Instructions
& \begin{tabular}[c]{@{}l@{}}
\( \displaystyle
dst \leftarrow *(src + off)
\)\\ \\ \\ \\ \\ \\
\( \displaystyle
dst \leftarrow *(src + off)
\)\\ \\ \\ \\ 
\( \displaystyle
dst \leftarrow *(src + off)
\)\\
\end{tabular}

% ################### LOAD Conditions
& \begin{tabular}[c]{@{}l@{}} 
\( \displaystyle
\regsrc.tag == \customvalue{pkt\_buff}, 
\)\\
\( \displaystyle
off == \text{getBuffOffset}_{spec}( 
\)\\ 
\( \displaystyle
% \custvariable{buff\_type},\ \customvalue{data}\ |\ \customvalue{data\_end}\
\custvariable{buff\_type},
\customvalue{data}\ |\ \customvalue{data\_end}\ 
% |\ \customvalue{<other\_field>})
\)\\
\( \displaystyle
|\ \customvalue{<other\_field>})
\)
\\ \\ \\ 
\( \displaystyle
\regsrc.tag == \customvalue{pkt\_data\_start}
\)\\ \\ \\ \\ 
\( \displaystyle
\regsrc.tag == \customvalue{stack\_frame}
\)\\
\end{tabular}

% ################### LOAD reg and mem states.
& \begin{tabular}[c]{@{}l@{}} 
% pkt_buff
\( \displaystyle
\regdst.tag \leftarrow \customvalue{pkt\_data\_start}\ 
\)\\
\( \displaystyle
|\ \customvalue{pkt\_data\_end}\ |\ \customvalue{pkt\_buff},
\)\\
\( \displaystyle
\regdst.val \leftarrow off
\)\\ \\ \\ 

% pkt_hdr
\( \displaystyle
\regdst.tag \leftarrow \regsrc.tag,
\)\\
\( \displaystyle
\regdst.val \leftarrow (\regsrc.val + off)
\)\\ \\ \\ 

% stack frame
\( \displaystyle
\regdst.tag \leftarrow \mem[off].tag,
\)\\
\( \displaystyle
\regdst.val \leftarrow \mem[off].val
\)
\end{tabular}

% ################### LOAD Context states
& \begin{tabular}[c]{@{}l@{}}
% pkt_buff
% \(\displaystyle
% \context{read\_buff}.tag \pluseq \customvalue{buff\_field},
% \)\\
\(\displaystyle
\context{read\_buff}.val \pluseq 
\)\\
\( \displaystyle
\text{getBuffFldName}_{spec}(buff\_type,\ off)
\) \\ \\ \\ \\

% pkt_start
% \(
% \displaystyle
% \context{read\_hdr}.tag \pluseq \customvalue{hdr\_field},
% \)\\
\( \displaystyle
\context{read\_hdr}.val \pluseq 
\)\\
\( \displaystyle
\text{getHdrFldName}_{spec}(\context{curr\_proto}.val,\ off)
\)
\\ \\ \\ 
% stack_frame
\multicolumn{1}{c}{\textbf{--}} \\
\end{tabular}
\\ \midrule

\begin{tabular}[c]{@{}l@{}} 
\textbf{R11:} Next \\ protocol check \\ \\ \\ 
\textbf{R12:} Memory \\ bound check 
% \textbf{R13:} Helper \\ function \\ invocation
\end{tabular}

% {}
%     { }
% ######### JUMP Instructions
& 
\begin{tabular}[c]{@{}l@{}} 
\( \displaystyle
dst \neq imm\ |\ dst == imm
\)\\
\\ \\ \\ 

\( \displaystyle
dst < src
\) 
% \( \displaystyle
% CALL\ imm
% \) \\
\end{tabular}

% ######### JUMP Conditions
& 
\begin{tabular}[c]{@{}l@{}} 
\( \displaystyle
\regdst.tag == \customvalue{pkt\_data\_start},
\)\\
\( \displaystyle
\text{isTailField}_{spec}(
\)\\
\( \displaystyle
\context{curr\_proto}.val,\regdst.val)
\) \\ \\ 

\(\displaystyle
\regdst.tag == \customvalue{pkt\_data\_start},
\)\\
\(\displaystyle
\regsrc.tag == \customvalue{pkt\_data\_end}
\)
\\ 
% \multicolumn{1}{c}{\textbf{--}} 
\end{tabular}

% ######### JUMP Reg. and Mem. state

& \multicolumn{1}{c}{
    \begin{tabular}[c]{@{}c@{}}  
        \textbf{--}\\ \\ \\ \\ 
        \textbf{--}\\
        % \textbf{--}
        \end{tabular}
} 

% ######### JUMP Context state
&
\begin{tabular}[c]{@{}l@{}} 
% \(\displaystyle
% \context{next\_proto\_check}.tag \leftarrow \customvalue{next\_proto\_check}, 
% \)\\
\(\displaystyle
\context{next\_proto}.val \leftarrow \text{getProtoName}_{spec}(
\)\\
\(\displaystyle
\custvariable{imm})
\)
\\ \\ 

% \(\displaystyle
% \context{next\_proto\_check}.tag \leftarrow \customvalue{bound\_check}, 
% \)\\
% \(\displaystyle
% \context{curr\_proto}.tag \leftarrow \customvalue{curr\_proto},
% \)
\\
\(\displaystyle
 \context{curr\_proto}.val \leftarrow \context{next\_proto}.val
\)

% \( \displaystyle
% \context{helper}.val \pluseq \text{getHlprFuncName}_{spec}(\custvariable{imm})
% \)
\end{tabular}
\\ \midrule

\begin{tabular}[c]{@{}l@{}} 
\textbf{R13:} Helper \\ function \\ invocation
\end{tabular}
&
\( \displaystyle
CALL\ imm
\)
&
\multicolumn{1}{c}{\textbf{--}} 
&
\multicolumn{1}{c}{\textbf{--}} 
&
\begin{tabular}[c]{@{}l@{}} 
\( \displaystyle
\context{helper}.val \pluseq \text{getHlprFuncName}_{spec}(\custvariable{imm})
\)
\end{tabular}

\\\midrule

% #### Packet actions (NF Return values)
\begin{tabular}[c]{@{}l@{}} 
\textbf{R14}: Return action
\end{tabular}
&
\( \displaystyle
return\ r_{0}
\)

& 
\multicolumn{1}{c}{\textbf{--}} 
& 
\multicolumn{1}{c}{\textbf{--}} 
& 

\eat{\begin{tabular}[c]{@{}l@{}}
% \( \displaystyle
% \context{pkt\_action}.tag \pluseq \mathbf{pkt\_action}
% \)
% \\
\( \displaystyle
\context{pkt\_action}.val \pluseq [hook,
\)\\
\( \displaystyle
\text{getActionName}_{spec}(\mathcal{R}[0].val)]
\)
\end{tabular}}

\begin{tabular}[c]{@{}l@{}}
% \( \displaystyle
% \context{pkt\_action}.tag \pluseq \mathbf{pkt\_action}
% \)
% \\
\( \displaystyle
\context{pkt\_action}.val \pluseq [hook,
\)\\
\( \displaystyle
\text{getActionName}_{spec}(\mathcal{R}[0].val),
\)\\
\( \displaystyle
\text{getPathAndNwContext()}]
\)
\end{tabular}
\\ \midrule

%%%%%%%%%%%%% Rules for Correlated maps %%%%%%%%%%%%%%%%%%%%%

\begin{tabular}[c]{@{}l@{}} 
\textbf{R15:} Correlated \\ maps
\end{tabular}
&
\( \displaystyle
CALL\ 1 | 2 | 3 | 51   
\)
% https://github.com/iovisor/bpf-docs/blob/master/eBPF.md
& 
\( \displaystyle
\mathcal{R}[2].tag == \customvalue{$\mathbf{Ref_{map}}$}
\)
& 
\multicolumn{1}{c}{\textbf{--}} 
& 

\eat{\begin{tabular}[c]{@{}l@{}}
% \( \displaystyle
% \context{pkt\_action}.tag \pluseq \mathbf{pkt\_action}
% \)
% \\
\( \displaystyle
\context{pkt\_action}.val \pluseq [hook,
\)\\
\( \displaystyle
\text{getActionName}_{spec}(\mathcal{R}[0].val)]
\)
\end{tabular}}

\begin{tabular}[c]{@{}l@{}}
\(\displaystyle
\context{correlated\_maps}.val \pluseq [\text{getMapName}_{elf}(),
\) \\
\(\displaystyle
\text{getMapName}_{elf()}]
\)
\end{tabular}
\\ \bottomrule

\end{tabular}
}

\end{table*}
% ####################################### transfer function rules END ################################## 

\section{\archp Analyzer} 
\label{sec:bca}

\myparab{Dataflow analysis for network context extraction.}
To extract network context, we depend on the widely used classic data flow analysis and propagation~\cite{hecht-book-77, kam1977monotone, KamUllman-dataflowAnalysis-76, kildall-framework-popl73} due to its scalability and comprehensive coverage of programs. 
Alternatives like SAT/SMT solvers~\cite{z3-smt} and abstract interpretation~\cite{cousot-cousot-1977-abstractinterpretation}, while being powerful for exhaustive state spaces and offer various benefits verifying program properties, come with significant limitations: primarily, they are computationally expensive~\cite{garey1979computers, chen23smt, cousot-halbwachs-1978}, and may not be well-suited for retrieving precise, detailed information about program behaviors, especially about bytecodes. 

In contrast, data flow analysis is well-suited for eBPF bytecode, which is simpler than traditional programs due to its lack of loops and fixed register count. This simplicity reduces the state space, making the analysis more scalable and manageable, where even complex NFs are handled with minimal overhead. Data flow analysis allows for granular insights into the network context at each program point, offering a more efficient and detailed approach.

\myparab{Challenges.} As discussed in \S\ref{sec:introduction}, two key challenges arise in extracting network context information at the eBPF bytecode level. This involves (1) interpreting instructions' semantics in relation to another, and (2) tracking dataflow information across register and memory locations involving redefinitions.

\myparab{Overview.} To address these challenges, we propose a formal flow analysis framework that integrates both register/memory state tracking and context state extraction.
Particularly, we define formal context state (C) rules that map low-level bytecode instructions to high-level program semantics. These rules capture important operations, such as reading packet headers or writing to maps, and protocol-specific behaviors, allowing us to accurately track how context evolves as the instructions are traversed (more details in~\tref{tab:inference_rules1} and ~\secref{subsec:nce}).
Also, we develop precise register (R) and memory (M) state rules to track the flow of data and interactions throughout the program. These rules handle the complexities of register (re-)definitions and memory updates, ensuring accurate tracking of data flow (more details in~\tref{tab:inference_rules1} and ~\secref{subsec:sm}).

Both the context state (C) and the register/memory state (R, M) are governed by a unified flow analysis framework. The register and memory states provide information for updating the context states. Context states are used to derive the bytecode's network context.

\subsection{\ControlFlowGraph}
\label{subsec:cfg_nc} 

\myparab{Basic block state and network context ($\mathcal{R}$, $\mathcal{M}$, and $\mathcal{C}$).} 
Our dataflow analysis also considers the control flow of the program, by making use of the Control Flow Graph. The absence of complex control flows in the eBPF code simplifies the control flow analysis.
We now explain the process of constructing a \ControlFlowGraph (\CFG) for a given bytecode. The CFG-NC has \textit{basic blocks} (BB) as nodes, and the edges capture the data propagated between two nodes. We group instructions into sequences that have a single entry point and a single exit point, ending either at a \textit{jump} instruction or just before a \textit{jump target}, and associate each such sequence with a basic block (BB).
The instructions in each BB are processed to generate two types of data: BB's program state ($\mathcal{R}$ and $\mathcal{M}$) and BB's network context $\mathcal{C}$. More specifically, the program state captures the state of registers ($\mathcal{R}$) and stack memory ($\mathcal{M}$) after processing a BB. The network context ($\mathcal{C}$) captures the BB's packet-processing behavior, such as read/write/copy operations on packet headers/buffers, maps accessed, and helper functions invoked. More details on how $\mathcal{R}$, $\mathcal{M}$, and $\mathcal{C}$ are generated in the following sections (\secref{subsec:sm} and \secref{subsec:nce}).

\myparab{State propagation and \CFG generation.} We design a forward analysis; $\mathcal{R}$, $\mathcal{M}$, and $\mathcal{C}$\footnote{We do not propagate the complete network context state, but only a portion that requires previous block's network context} state of a BB ($OUT[BB]$) forms the $IN$ state ($IN[BB]$) to all its successors in the CFG. A BB's $OUT$ state is computed as a \textit{function} of $IN$ state where the \textit{function} applies a set of state transformation rules (as shown in \tref{tab:inference_rules1}) on instructions in the BB.
More details on the instruction categories and the rules applied to each category are in Section \secref{subsec:nce}. 
If a BB has more than one predecessor, the $OUT$ state of such predecessor is merged using a \textit{confluence} operator and prepares $IN$ set. That is,

\begin{equation*}
    IN[BB] = \bigcup_{p \in pred(BB)} OUT[p]
\end{equation*}

The BBs in CFG are processed in depth-first search (DFS) order and the network context ($\mathcal{C}$) extracted from each BB is annotated. Subsequently, the query engine module answers user queries by processing $\mathcal{C}$ in the CFG-NC (\secref{sec:query_engine}).

\begin{figure}[tb!]
     \centering
     % \hspace{-0.7cm}
     \includegraphics [width=0.47\textwidth, keepaspectratio]{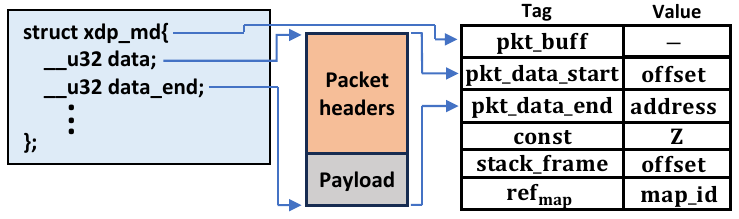}
     % \vspace{-0.6cm}
     \caption{R and M <tag,value> pairs. Z means integer.}
     \label{fig:R_M_states}
     \vspace{-0.4cm}
 \end{figure}

\begin{figure*}[t]
     \centering
     % \hspace{-0.7cm}
     % \includegraphics [width=19cm, keepaspectratio]{figures/context_extraction}
     \includegraphics [width=\linewidth, keepaspectratio]{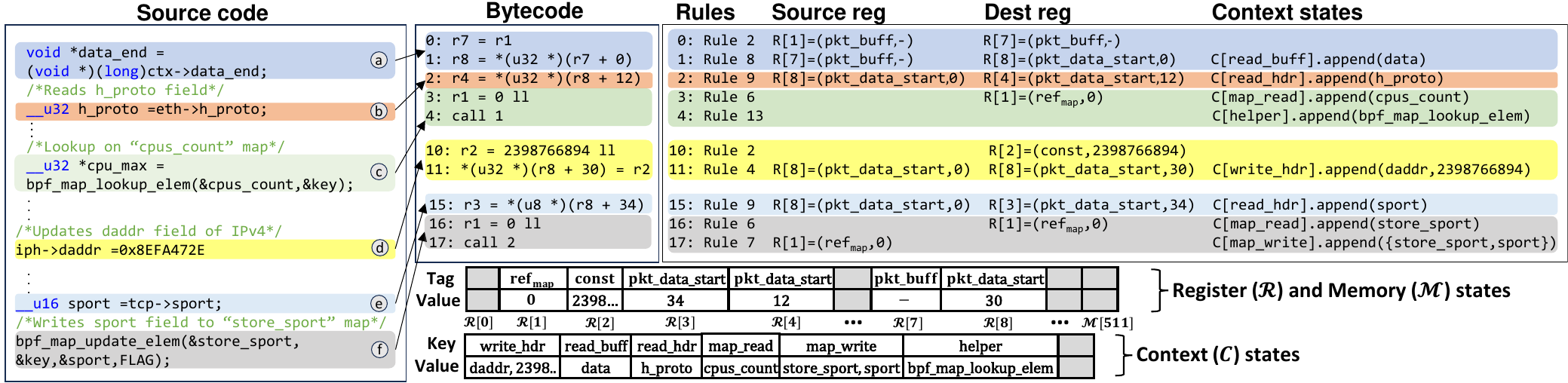}
     % \vspace{-0.6cm}
     \caption{Operation \textcircled{a} reads \textit{data} field from packet buffer. Operation \raisebox{.5pt}{\textcircled{\raisebox{-.9pt}{b}}} reads \textit{Ethertype} (h\_proto) field from ethernet header. Operation \textcircled{c} performs lookup on \textit{cpus\_count} map. Operation \raisebox{.5pt}{\textcircled{\raisebox{-.9pt}{d}}} updates the \textit{destination IP address} of IPv4 header. Operation \textcircled{e} reads \textit{source port} from TCP header. Operation \raisebox{.5pt}{\textcircled{\raisebox{-.9pt}{f}}} writes \textit{source port} to \textit{store\_sport } map.}
     
     \label{fig:running_example}
 \end{figure*}

\subsection{$\mathcal{R}$, $\mathcal{M}$, and $\mathcal{C}$ data structures}{\label{subsec:sm}}
\label{subsec:data_structure}

We present $\mathcal{R}$, $\mathcal{M}$, and $\mathcal{C}$ data structures followed by state transformation rules (\tref{tab:inference_rules1}) applied on instructions. 

\myparab{R and M data structures.}  Instructions in eBPF bytecode use up to $11$ registers and a program stack of $512$ bytes. Complex eBPF programs may run out of registers and often use the program stack whenever register spills are observed~\cite{classic_bpf_vs_ebpf}. Therefore, analyzing some instructions requires tracking and updating $\mathcal{R}$ and $\mathcal{M}$. 
$R\left[n\right ]$ represents register states where $n \in \{0,1,2, \dots,10\}$, one each for $11$ registers ($r_0$-$r_{11}$) used in instructions. $M\left[k\right ]$ represents $k^{th}$ byte of the program stack, where $k \in \{0,1,2,\dots,511\}$. Each entry $r_n$  and byte $k$ has a (\textit{tag}, \textit{value}) pair. Based on possible values, we came up with six tags as shown in~\fref{fig:R_M_states}: (i) tag \customvalue{pkt\_buff} indicates $r_n$ or byte $k$ holds a reference to a packet struct buffer (e.g., \textit{xdp\_md}, \textit{\_\_sk\_buff}) with value as none, (ii) \customvalue{pkt\_data\_start} contains an offset from the start of packet data, (iii) \customvalue{pkt\_data\_end} refers to the end of the packet data, (iv) \customvalue{const} holds the constant value, (v) \customvalue{$\mathbf{ref}_{\mathbf{map}}$} refers to a eBPF map in the program, and (vi) \customvalue{stack\_frame} refers to a location in the stack.

\myparab{Network Context ($\mathcal{C}$) data structure.} Similar to $\mathcal{R}$ and $\mathcal{M}$, $\mathcal{C}$ also has a list of (\textit{tag}, \textit{value}) pairs. There are seven tags: (i) \customvalue{hdr\_field} and \customvalue{buff\_field} capture the list of packet header and packet buffer fields accessed or updated; (ii) \customvalue{helper\_function} captures the list of helper functions invoked; (iii) \customvalue{map} captures the names of eBPF map accessed; and  (iv) [\customvalue{next\_proto}, \customvalue{bound\_check}, \customvalue{curr\_proto}], all three help to extract protocols processed in the bytecode (\S\ref{subsec:nce}).

\myparab{Initialization:} Conventionally, register $r1$ points to the packet buffer at the beginning of the eBPF bytecode -- packet buffer structures are defined in the kernel header \texttt{bpf.h} (see~\fref{fig:R_M_states}). These structures provide input context to the eBPF programs and hold the pointers to the start and end of the packet~\cite{fast_packet_proce_ebpf_xdp}. Hence, the state of register $r1$ is initialized to (\textit{pkt\_buff}, -), and other states ($\mathcal{R}$, $\mathcal{M}$, and $\mathcal{C}$) are set to NULL.

\subsection{Network context extraction}{\label{subsec:nce}}

\myparab{State transformation rules by instruction category.}  We apply transfer function rules to (i) \textit{propagate} $\mathcal{R}$ and $\mathcal{M}$ states from one instruction to another within and across BBs, and (ii) \textit{record} $\mathcal{C}$ in each BB. An instruction in bytecode~\cite{ebpf_spec} contains opcode ($8$ bits), source and destination register ($4$ bits each), offset ($16$ bits), and immediate value ($32$ bits). We divide instructions into $8$ categories, as shown in \tref{tab:inference_rules1}, and apply associated rules. Each rule has an optional condition, and if the condition is satisfied, $R$ and $M$ states are updated, and the associated network context is recorded in $C$. Variables are italicized (e.g., $\mathcal{R}[src].tag$, $\mathcal{R}\left[src\right].val$, \textit{imm}, etc.) and the tags are specified in bold (e.g., \textbf{pkt\_buff}, \textbf{pkt\_data\_start}).

\myparab{Protocol information and support functions.} Bytecode instructions contain immediate values and offsets, each with a specific meaning. Such as field offsets in packet headers or packet buffers (\eg, sk\_buff), protocol number (\eg, ethrType, ipv4.protocol), helper function ID, action ID specific to a hook point, and map ID. We capture these IDs in a file and define a set of support functions that perform look-up on the file and return names (\eg, protocol name, helper function name) that an end user can understand. It can be extended to support new protocols, hook points, and helpers.

\myparab{Running examples.} For simple arithmetic (R1) and assignment (R2) operations, we update R and M states, but do not record network context because these instructions would not directly update/read packet headers/buffers. In contrast, we record network context for write operations (R3-R5) and read operations (R8-R10). For instance, in~\fref{fig:running_example}, we show bytecode instructions $1$, $2$, and $15$ are read operations and transfer function rules R8, R9, and R9 are applied, respectively\footnote{`getBuffOffset' support function get the offset to a packet buffer field as defined in the spec.}. Instruction 11 is of write type, so rule R4 is applied\footnote{`getHdrFldName' and `getBuffFldName' support functions return the names of the header field and packet buffer, respectively.}.

R6, R7, and R13 are applied for map lookup, map update, and helper functions, respectively. For instance, consider bytecode instructions $3$ and $4$ in~\fref{fig:running_example}. These instructions call the lookup helper function on the $cpus\_count$ map. On these instructions, R6 and R13 are applied to extract the names of the map and the helper function invoked. Similarly, byte instructions $16$ and $17$ calls map update helper function on $store\_sport$ map. For these instructions, we record the names of maps, fields written to map, and the helper functions associated with the \textit{imm} value\footnote{`getMapName\textsubscript{elf}()' does a look-up on ELF section of the object file and returns map's name.}.

Rules R11 and R12 are applied to extract the list of protocol headers processed in the bytecode. Generally, while parsing packet data, a header is identified by looking at specific fields in the previously identified headers. For example, consider the first 14 bytes as an Ethernet header. Then, looking at the etherType field, we know the next header is a VLAN, an IPv4, or an IPv6 header. Similarly, looking at the ipv4.proto field, we know the next header is a TCP or a UDP header.
We leverage the packet header semantics~\cite{panda-parser} and recognize headers. We also leverage that most bytecodes perform memory-bound checks before accessing the next header to ensure subsequent code accesses within the packet memory. If the next header is accessed without a memory-bound check, then the eBPF verifier throws an error and will not allow the loading of the bytecode. A running example bytecode with steps to illustrate this is in~\fref{fig:proto_running_example} in Appendix~\ref{app:hid}.

R13 retrieves the eBPF helper function invoked based on \textit{imm} value, which represents an ID associated with each helper function.
Helper function names are retrieved using the `getHlprFuncName' support function. R14 captures the return action (\eg, XDP\_PASS, TC\_ACT\_OK) based on register $r_{0}$ value, which varies for different \textit{hook} points\footnote{`getActionName' support function retrieves the action name.}(\eg, XDP, TC). It also appends the current path's network context to $\mathcal{C}$. Two maps, say A and B, are considered correlated if map A's lookup result is map B's key.  Rule R15 retrieves the correlated maps based on register $r_{2}$ tag; it holds the key for map operations such as map update (call 2) and delete (call 3).

%\vspace{-5mm}
\section{Language and Query Engine (QE)} 
\label{sec:query_engine}

\myparab{\archp Language.} The grammar, as shown in \fref{fig:grammar}, follows a declarative (logical) program paradigm~\cite{prolog_book} with syntax similar to the widely used Prolog\footnote{Prolog is just an implementation language used by \archp; other logic programming languages could also be used.} language syntax~\cite{the_art_of_prolog, the_craft_of_prolog, swi_prolog}. The grammar exposes specialized eBPF-specific predicates that enable eBPF users, the primary target audience of the tool, to express assertions and queries easily by leveraging their existing knowledge of eBPF.
The two key constructs are: (1) \textbf{\qpredslong}: \qpredslong (\qpred) are the primary building blocks of queries. They assert specific properties or retrieve network context. \qpred takes one or more arguments: when all arguments of a \qpred are defined with a specific value, the predicate asserts that the property holds. When one or more arguments are variables (\ie, not defined), the \qpred retrieves all instances where the property applies, with variables being filled by the retrieved values, and (2) \textbf{Queries}: A query consists of one or more \qpreds connected by logical AND(,)/OR(;) operator and ends with period (.).

% ############################## Table for Mapping facts to Network context collected:
\begin{table}[t]
\centering
\scriptsize
\caption{Predicates, network context, and associated facts.}
\label{tab:context_to_facts_mapping}
\begin{tabular}{p{0.125\linewidth} p{0.35\linewidth} p{0.39\linewidth}}
\toprule

\textbf{\begin{tabular}[c]{@{}l@{}}Predicate\\Category\end{tabular}} 
& \textbf{\begin{tabular}[c]{@{}l@{}}Network Context from\\\CFG\end{tabular}} 
& \multicolumn{1}{c}{\textbf{Facts}}                                                             
\\ \toprule

\begin{tabular}[c]{@{}l@{}}Field\\predicates\end{tabular}              
& \begin{tabular}[c]{@{}l@{}}$\mathcal{C}${[}\textit{read\_buff}{]} = {[}(buff\_field,\\\rightline{field\_name)],}\\ $\mathcal{C}${[}\textit{read\_hdr}{]} = {[}(hdr\_field,\\\rightline{field\_name)],}\\ $\mathcal{C}${[}\textit{write\_buff}{]} = {[}(buff\_field,\\\rightline{field\_name)],}\\ $\mathcal{C}${[}\textit{write\_hdr}{]} = {[}(hdr\_field,\\\rightline{field\_name)]}\end{tabular} 
& \begin{tabular}[c]{@{}l@{}}read\_buffer\_field("nf\_id", "bb\_id",\\\rightline{field\_name),}\\read\_header\_field("nf\_id", "bb\_id",\\\rightline{field\_name),}\\ write\_buffer\_field("nf\_id", "bb\_id",\\\rightline{field\_name),}\\ write\_header\_field("nf\_id", "bb\_id",\\field\_name)\end{tabular}
\\ \midrule

\eat{\begin{tabular}[c]{@{}l@{}}Map\\predicates\end{tabular}                
& \begin{tabular}[c]{@{}l@{}}$\mathcal{C}${[}\textit{read\_map}{]} = {[}(map,\\\rightline{map\_name)],}\\ $\mathcal{C}${[}\textit{write\_map}{]} = {[}(map,\\\rightline{map\_name, field\_name)]}\end{tabular}
& \begin{tabular}[c]{@{}l@{}}read\_from\_map("nf\_id", "bb\_id",\\\rightline{map\_name),}\\ write\_into\_map("nf\_id", "bb\_id",\\\rightline{map\_name, field\_name)}\end{tabular}                                                                                                  
\\ \midrule}

\begin{tabular}[c]{@{}l@{}}Map\\predicates\end{tabular}                
& \begin{tabular}[c]{@{}l@{}}$\mathcal{C}${[}\textit{read\_map}{]} = {[}(map,\\\rightline{map\_name)],}\\ $\mathcal{C}${[}\textit{write\_map}{]} = {[}(map,\\\rightline{map\_name, field\_name)]}\\ $\mathcal{C}${[}\textit{correlated\_maps}{]} = {[}(map,\\\rightline{map\_A, map\_B)]}
\end{tabular}
& \begin{tabular}[c]{@{}l@{}}read\_from\_map("nf\_id", "bb\_id",\\\rightline{map\_name),}\\ write\_into\_map("nf\_id", "bb\_id",\\\rightline{map\_name, field\_name)}\\ correlated\_maps("nf\_id", "bb\_id",\\\rightline{map\_A, map\_B)}
\end{tabular}                                                                                                  
\\ \midrule

\begin{tabular}[c]{@{}l@{}}Helper\\predicate\end{tabular}             
& \begin{tabular}[c]{@{}l@{}}$\mathcal{C}${[}\textit{helper}{]} = {[}(helper,\\\rightline{function\_name)]}\end{tabular}                                                                                               
& \begin{tabular}[c]{@{}l@{}}invoke\_helper("nf\_id", "bb\_id",\\\rightline{function\_name)}\end{tabular}
\\ \midrule

\begin{tabular}[c]{@{}l@{}}Protocol\\predicate\end{tabular}          
& \begin{tabular}[c]{@{}l@{}}$\mathcal{C}${[}\textit{curr\_proto}{]} = [(curr\_proto,\\\rightline{protocol\_name)]}\end{tabular}
& \begin{tabular}[c]{@{}l@{}}protocol\_accessed("nf\_id", "bb\_id",\\\rightline{protocol\_name)}\end{tabular}
\\ \midrule

\eat{\begin{tabular}[c]{@{}l@{}}Packet action\\predicate\end{tabular}     
& \begin{tabular}[c]{@{}l@{}}$\mathcal{C}${[}\textit{pkt\_action}{]} = {[}(hook,\\\rightline{action\_name)]}\end{tabular}
& \begin{tabular}[c]{@{}l@{}}return\_action("nf\_id", "bb\_id",\\\rightline{hook, action)}\end{tabular}
\\ \midrule}

\begin{tabular}[c]{@{}l@{}}Action\\predicate\end{tabular}     
& \begin{tabular}[c]{@{}l@{}}$\mathcal{C}${[}\textit{pkt\_action}{]} = {[}(hook, \\\rightline{action, $P$)]}\end{tabular}
& \begin{tabular}[c]{@{}l@{}}return\_action("nf\_id", hook, \\\rightline{action, $P$)}\end{tabular}
\\ \midrule

\begin{tabular}[c]{@{}l@{}}Order\\predicates\end{tabular}             
& \multicolumn{1}{c}{\textbf{--}} 
& \begin{tabular}[c]{@{}l@{}}edge("bb\_id", "bb\_id"),\\nf\_edge("nf\_id", "nf\_id")\end{tabular}
\\ \bottomrule
\end{tabular}

\end{table}

\subsection{Query Engine}
\label{sub:answering_queries}

Query Engine (QE) translates network context encoded in \CFG to prolog facts while maintaining the structural integrity of the \CFG and order of NFs in the NF chain. The user queries written using \archp language (\fref{fig:grammar}) are answered by matching \qpreds against the facts in the knowledge base. Table~\ref{tab:context_to_facts_mapping} illustrates \qpredslong\footnote{definitions are in Appendix~\ref{app:predicates}}, prolog facts, and corresponding network context in \CFG. Facts are generated from the network context associated with each \CFG node. We call the complete set of facts for all bytecode(s) in an NF chain the knowledge base.

\begin{figure}[t]
     \includegraphics [width=8.5cm]{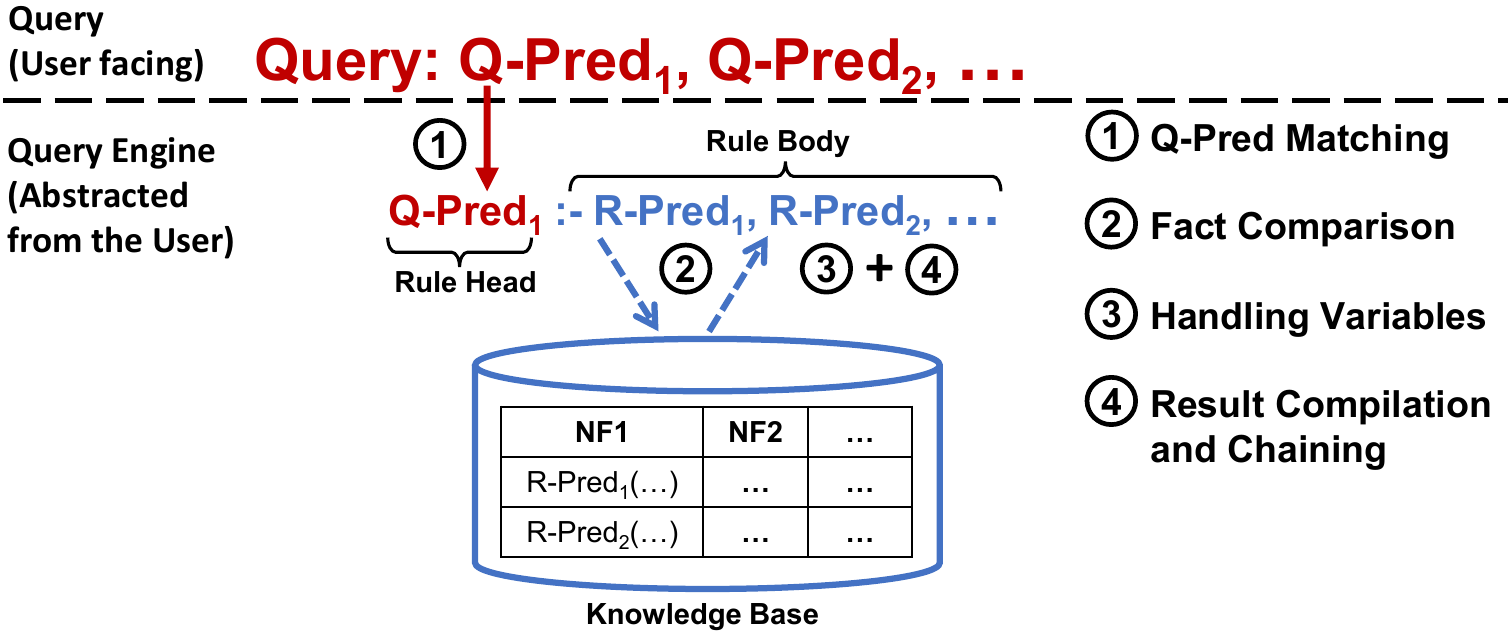}
    \caption{Query execution workflow.} 
     \label{fig:query_work_flow}
\vspace{-0.6cm}
 \end{figure}

For mapping \qpreds to facts, internally, we define rules that help in matching. The rules are of the form \texttt{\qpred}:- \texttt{\rpred}\textsubscript{\texttt{1}}\textit{op} \texttt{\rpred}\textsubscript{\texttt{2}}\textit{op} \texttt{\rpred}\textsubscript{\texttt{3}}\textit{op} $\dots$, \texttt{\rpred}\textsubscript{\texttt{n}}, where \texttt{\qpred} is the query predicate which constitutes the head of the rule, \texttt{\rpreds} correspond to the rule predicates constituting the rule body and \textit{op} are the logical operators such as AND(,) and OR(;). As shown in \fref{fig:query_work_flow}, rule heads are exposed to the operator (\S\ref{fig:grammar}), and the body constitutes the operational semantics of the particular query predicate. This approach abstracts the details of the underlying facts while allowing users to write queries without needing to understand the detailed structure of the underlying facts. Table~\ref{tab:prolog_rules} in Appendix~\ref{app:opsemantcis} summarizes the operational semantics of all \qpreds.

The following steps are involved in answering the user queries from the knowledge base. (1) \textbf{\qpred matching:} Each \qpredlong in a query is compared against the heads of the rules defined. (2) \myparab{Fact comparison:} The \rpreds are matched against the facts stored in the knowledge base. (3) \myparab{Handling variables:} If the query contains variables, the engine retrieves all values that satisfy the rule from the knowledge base. (4) \myparab{Result compilation and chaining:} After the matching process, results are compiled by collecting all the values that satisfy the query predicates. Appendix~\ref{app:kbase} presents a running example explaining these steps.

\myparab{\archp assertion execution.} In~\secref{subsec:AE}, we explain the workings of Analyzer and Query Engine using three assertions, \textbf{A1}, \textbf{A2} and \textbf{A3} defined in Sec.~\secref{subsec:motivating_examples}, on two bytecodes (Listing~\ref{lst:firewall_snippet} and~\ref{lst:update_mark} in Sec.~\secref{subsec:motivating_examples}).

\myparab{Extensions.} \archp can be extended to support new user-defined \qpredslong to match the facts.
As discussed, the new \qpredslong should have semantics with the corresponding \rpreds, and they can be incorporated as a library to support more queries for new use cases.

\section{Evaluation}
\label{sec:evaluation}

We evaluate \archp to study: 1) Expressiveness of \archp language, 2) Usage of \arch-Prashna (\secref{subsec:usage_of_Yaksha}), 3) Accuracy of network context extraction (\S\ref{subsec:net_cont_extract}), 4) time to generate \CFG (\S\ref{subsec:net_cont_extract}) and to execute queries (\S\ref{subsec:query_exec}), and 5) Memory overheads (\S\ref{subsec:mem_footprint}).

\myparab{Implementation.}  
\archp analyzer module consists of $1.8$K LoC in C++. Internally, QE module invokes a SWI-Prolog (version 9.3.3)~\cite{swi_prolog} environment, which takes facts and rules and provides a CLI for writing queries. This module consists of $~600$ LoC in C++. We ran experiments on a server with AMD EPYC 7262, 3.2 GHz, 8-core CPU, 32GB RAM.

% ###########################################
\begin{table*}[]
\caption{Properties supported by \archp, Klint and DRACO. (\protect\halftick): partial support. Property's superscript has NF name -- RT: Router, FW: Firewall, LB: Load balancer, NAT: NAT, BR: Bridge, POL: Policer, KTRN: Katran LB, CRAB: CRAB LB, FLU: Fluvia. %hXDP-FW: Firewall from hXDP. 
$*$ represents all NFs}
\label{tab:comparison}
\begin{tabular}{|l|l|c|c|c|}
% \begin{tabular}{llccc}
\hline
\textbf{Category}				         
&	\textbf{Properties(P1-P22)/Retrieval Queries(Q23-Q24)}
&	\textbf{\archp}                  
&	\textbf{Klint} 
&	\textbf{DRACO} \\ \hline

&	P1. Expected header$^{(*)}$ 	&	\checkmark	&	\checkmark	&	-	\\ \cline{2-5}	
&	P2. Validity of IPv4 header$^{(RT)}$	&	\checkmark	&	\checkmark	&	-	\\ \cline{2-5} 
&	\begin{tabular}[c]{@{}l@{}}P3. Expected header field updates by NF$^{(RT)}$\end{tabular}	&	\checkmark	&	\checkmark	&	-	\\ \cline{2-5} 
&	P4. Longest prefix matching$^{(RT)}$	&	\halftick	&	\checkmark	&	-	\\ \cline{2-5} 
&	P5. Packet content preservation$^{(FW)(LB)}$	&	\checkmark	&	\checkmark	&	\checkmark	\\ \cline{2-5} 
&	P6. Filtering external flows$^{(FW)(NAT)}$	&	\checkmark	&	\checkmark	&	-	\\ \cline{2-5} 
&	P7. Remembering internal flows$^{(FW)(NAT)}$	&	\checkmark	&	\checkmark	&	-	\\ \cline{2-5} 
&	\begin{tabular}[c]{@{}l@{}}P8. Update the source IP and port of the\\ internal flow$^{(NAT)}$\end{tabular}	&	\checkmark	&	\checkmark	&	-	\\ \cline{2-5} 
&	\begin{tabular}[c]{@{}l@{}}P9. Restore the destination IP and port of\\ the external flow$^{(NAT)}$\end{tabular}	&	\checkmark	&	\checkmark	&	-	\\ \cline{2-5} 
&	\begin{tabular}[c]{@{}l@{}}P10. Selected transmission port must not be\\ the ingress port$^{(BR)}$\end{tabular}   	&	\checkmark	&	\checkmark	&	-	\\ \cline{2-5} 
&	P11. Broadcast if unknown$^{(BR)}$	&	\halftick	&	\checkmark	&	-	\\ \cline{2-5} 
&	P12. The learning process$^{(BR)}$	&	\checkmark	      	&	\checkmark	&	-	\\ \cline{2-5} 
&	P13. External traffic policing$^{(POL)}$	&	\checkmark	&	\checkmark	&	-	\\ \cline{2-5} 
&	P14. Backend liveness$^{(LB)}$	&	\checkmark	&	\checkmark	&	-	\\ \cline{2-5} 
&	P15. Load balancing external traffic$^{(LB)}$	&	\checkmark	&	\checkmark	&	-	\\ \cline{2-5} 
&	P16. Drop all fragmented packets$^{(KTRN)}$	&	\checkmark	&	\checkmark	&	\checkmark	\\ \cline{2-5} 
&	\begin{tabular}[c]{@{}l@{}}P17. Forward all ICMP echo request\\ packets$^{(KTRN)}$\end{tabular}	&	\checkmark	&	\checkmark	&	\checkmark	\\ \cline{2-5} 
&	P18. Handle only TCP SYN packets$^{(CRAB)}$	&	\checkmark	&	\checkmark	&	\checkmark	\\ \cline{2-5} 
&	P19. Add custom redirection header$^{(CRAB)}$	&	\checkmark	&	\checkmark	&	\checkmark	\\ \cline{2-5} 
&	P20. Forward all packets unchanged$^{(FLU)}$	&	\checkmark	&	\checkmark	&	\checkmark	\\ \cline{2-5} 
\multirow{-21}{*}{\begin{tabular}[c]{@{}l@{}}Category 1:  Individual NF \\ conformance to specification\end{tabular}}	&	P21. Correlated maps$^{(hXDP-FW)}$	&	\checkmark	&	\checkmark	&	\checkmark	\\ \hline

\begin{tabular}[c]{@{}l@{}}Category 2:  Unintended NF \\ interactions\end{tabular}	& \begin{tabular}[c]{@{}l@{}}P22. NF dependency: Read after \\ write (RAW), Write after read (WAR) \\ and Write after write (WAW)\end{tabular}     & \checkmark	& \crossmark	& \checkmark	\\ \hline

& \begin{tabular}[c]{@{}l@{}}Q23. Find the list of packet fields \\ updated by the NF\end{tabular}	& \checkmark	& \crossmark	& \crossmark	\\ \cline{2-5} 

\multirow{-4}{*}{\begin{tabular}[c]{@{}l@{}}Category 3: Understanding \\the NF/NF chain behavior \end{tabular}}	& \begin{tabular}[c]{@{}l@{}}Q24. Find the list of buffer fields updated\\ by one NF and read by successor NF \\(RAW dependent)\end{tabular} & \checkmark	& \crossmark	& \crossmark	\\ \hline

\end{tabular}
\end{table*}

\vspace{-2mm}
\subsection{Expressiveness}\label{subsec:expressiveness}
We evaluate language expressiveness by expressing properties and queries covering three categories~\secref{subsec:motivating_examples} over half a dozen standard NFs, four non-standard but real eBPF-based NFs and NF chains. \tref{tab:comparison} lists the properties supported by \archp, and closely related work Klint~\cite{verify_nf_binaries} and DRACO~\cite{draco_lu2024towards}. Each property is expressed in \archp's language (\tref{tab:literature-queries-new}), with a detailed explanation in Appendix \secref{subsec:comp_overview}. In this section, we summarize key observations.

The first category has $21$ properties (P1-P21), as shown in~\tref{tab:comparison}, which are derived from the standards documents (RFCs and IEEE standards) for individual NFs. \archp and Klint support most properties, whereas DRACO's support is unclear ($-$), except for Property P5 and Properties P16–P21, as it does not disclose the specification language. Properties P4 and P11 are partially supported by \arch-Prashna, because the language lacks predicates on map type and broadcast packets. The second category has properties on the NF chain to detect unintended interactions (\secref{subsec:motivating_examples}). Currently, Klint supports assertions only on individual NFs, whereas \archp and DRACO support assertions on dependencies in an NF chain.

The third category has retrieval queries (Q23 and Q24) that demonstrate the language's ability to support not only assertions, but also to retrieve specific details causing violations. More specifically, to find which field was modified by a firewall (Q23) that led to an assertion violation,  Klint or DRACO would require writing a preservation assertion for each header field, such as \textit{“Is the Source MAC modified?”} or \textit{“Is the Source IP modified?”}, etc., which is tedious or infeasible as it requires checking every possible header and takes a long time for large programs. In comparison, \archp can get this information instantly from the facts stored in the knowledge base. Similarly, \archp can also retrieve the buffer fields (Q24), causing RAW dependency (Listing \ref{lst:update_mark}).

% ############################################################

\begin{table}[t]
% \scriptsize
\centering
\caption{Microbenchmark NFs. (*) NFs from open source projects, (+) Hand-written NFs, NI: \# Instructions, NAI: \#Instructions analyzed, and NP: \# Paths.}
    \label{tab:microbenchmark}
\resizebox{1\columnwidth}{!}{
\begin{tabular}[t]{lllrrr}
\toprule
\multicolumn{1}{l}{\textbf{ID}}
&
\textbf{Network Functions}                 & \textbf{Project}                   & \textbf{NI}          &  \textbf{NAI}     & \textbf{NP}
\\ \midrule
NF1
&
\texttt{xdp\_pktcntr}   & Katran~\cite{katran}                                                              & 22                  & 28                    & 4                                                                                                      \\ \midrule

NF2
&
\texttt{xdp\_map\_access}    &       K2~\cite{k2}                                                & 42                  & 64                 &  6                                                                
    \\ \midrule

NF3
&
\texttt{storeIPinMap+}     &         \texttt{--}                                                & 39                  & 49                   & 7                                                                                                      \\ \midrule

NF4
&
\texttt{xdp\_fw}          &     K2~\cite{k2}                                                 & 73                  &  496                &   38                                                                                                    \\ \midrule

NF5
&
\begin{tabular}[c]{@{}l@{}}\texttt{ratelimiting\_kern\_new}\end{tabular}      &  ebpf-ratelimiter~\cite{ebpf-ratelimiter}                                           & 136                  & 292                 & 19                                                                                                      \\ \midrule

NF6
&
\texttt{ratelimiting\_kern}          & ebpf-ratelimiter~\cite{ebpf-ratelimiter}                                              & 167                  & 278                  & 18                                                                                                      \\ \midrule

NF7
&
\texttt{xdp1\_kern}        &    K2~\cite{k2}                                                   & 61                  &  925                &  99                                                                                                     \\ \midrule

NF8
&
\texttt{decap\_kern}        &   Katran~\cite{katran}                                                       & 228                  & 430                  & 33                                                                                                      \\ \midrule

NF9
&
\texttt{xdp\_filter}      &     Suricata~\cite{suricata}                                                      & 344                  & 13K                  & 639                                                                                                      \\ \midrule

NF10
&
\texttt{xdp\_filter\_buff\_update*}     &     Suricata~\cite{suricata}                                             & 348                  & 13K                  & 639                                                                                                     \\ \midrule

NF11
&
\texttt{mptm\_encap\_xdp}           &    Mptm~\cite{mptm}                                           & 348                  & 25K                 & 1K                                                                                                      \\ \midrule

NF12
&
\texttt{mptm\_decap\_xdp}           &       Mptm~\cite{mptm}                                         & 149                  & 38K                  & 2K                                                                                                      \\ \midrule

NF13
&
\texttt{xdp\_lb}              &          Suricata~\cite{suricata}                                           & 419                  & 65K                 & 4K                                                                                                      \\ \midrule

NF14
&
\begin{tabular}[c]{@{}l@{}}\texttt{xdp\_lb\_hdr\_update*}\end{tabular}       & Suricata~\cite{suricata}                                              & 423                  &  65K                & 4K                                                                                                      \\ \midrule

NF15
&
\begin{tabular}[c]{@{}l@{}}\texttt{xdp\_lb\_syn\_storeIPinMap*}\end{tabular}    &   Suricata~\cite{suricata}  & 440                  &  66K                & 4K                                                                                                      \\\midrule

NF16
&
\begin{tabular}[c]{@{}l@{}}\texttt{xdp\_lb\_syn\_storeIPTTLinMap*}\end{tabular}     &   Suricata~\cite{suricata}    & 460                  &  68K                & 4K                                                                                                      \\
\bottomrule
% \vspace{-20pt}
    \end{tabular}%
}
\end{table}

% #############################

\subsection{Usage of \archp}
\label{subsec:usage_of_Yaksha}
We demonstrate on a testbed server how an operator can use \archp to check whether an NF bytecode conforms to its specification and if not, why it deviates from expected NF behavior (Category 1). Also, it captures unintended NF interactions (Category 2), and aids in understanding NF/NF-chain behavior (Category 3), as motivated in \secref{subsec:motivating_examples}.

\myparab{Category 1 use case.} Consider that an NF operator wants to deploy a stateful firewall (Listing~\ref{lst:firewall_snippet}) with two bugs at the server’s XDP hookpoint via bpfman (see~\fref{fig:bpfman-workflow} in~\secref{app:bpfman_deployment}). The two bugs are: (1) updates the TCP source port (line 10), and (2) allows ICMP traffic (line 6). Before deploying, the operator uses \archp to verify against two assertions, A1 and A2.
A1 asserts that the firewall should not modify packet contents. Since the firewall modified the TCP source port, A1 failed, and \archp raised an alert. To understand the reason, the operator runs a retrieval query RQ1 (Category 3), which correctly identifies the violation source as TCP source port update. Similarly, A2 asserts that ICMP traffic should be dropped. Since the firewall allowed ICMP packets, A2 failed, and \archp raised another alert. Thus, through these assertions, \archp successfully detected and prevented deployment of a non-standard NF.

\myparab{Category 2 use case.} We emulate a real-world RAW dependency between Cilium and the AWS Elastic Network Interface. An NF chain of three XDP-based NFs (Listing~\ref{lst:update_mark}) was considered, with NF1 and NF3 deployed in sequence (NF1 → NF3). Before inserting NF2, the bytecodes of intended ordering (NF1 → NF2 → NF3) were verified using Assertion A3, which asserts that NF2’s write set does not overlap with NF3’s read set. Assertion A3 failed due to an overlap on \textit{sk\_buff->mark}, and \archp raised an alert. Next, using retrieval query RQ3 (Category 3), we pinpoint the cause of the violation. Hence, \archp prevented an unintended interaction that would have occurred upon deployment of a new NF into the existing NF chain.

\subsection{Network context extraction}
\label{subsec:net_cont_extract}

\myparab{NFs and Queries.} We generate bytecodes using \textit{LLVM Clang} (V11.1.0) with O2, Oz, and O3 flags for a microbenchmark suite of sixteen XDP-based NFs (\tref{tab:microbenchmark}). 
The selected NFs represent a diverse set of real-world XDP programs drawn from widely used open-source and production systems. They cover common packet-processing tasks, including packet counting, map access, firewalling, rate limiting, and load balancing. The suite spans a broad range of complexities and instruction counts: simple NFs (e.g., NF1 and NF2) implement basic functionality such as packet counting with 22–42 instructions, whereas more complex NFs (e.g., NF4, NF9, and NF13) perform firewalling, filtering, and load balancing with several hundred instructions, with the largest exceeding 400. This diversity enables us to evaluate \oursystem\ across both lightweight and compute-intensive NFs representative of practical deployments.
We design a total of $16$ queries (Table~\ref{tab:queries-eval} in Appendix~\ref{app:queries}) and assess the structural properties of individual NFs and the interactions across multiple NFs in an NF-Chain. These queries are evenly divided, with $8$ focusing on assertions and $8$ on retrieval, covering all predicate and rule categories detailed in~\tref{tab:context_to_facts_mapping} and~\tref{tab:prolog_rules}.

\myparab{Accuracy.} We define accuracy as the proportion of network context information extracted by the analyzer compared to the information in the ground truth. To establish the ground truth, we manually review the source code to extract list of protocols, specific fields accessed in each protocol/buffers, helper functions, and maps. The analyzer extracts information correctly for all NFs, except for some fields in NF12. We also compare the assertion query results (\ie, Q1(A)--Q8(A)) for randomly stacked facts of microbenchmark NFs (\tref{tab:microbenchmark}) and found the answers are correct.

\Oursystem\ assumes that an NF performs a next-protocol check followed by a memory-bound check before accessing protocol header fields. This assumption is derived from real-world implementations and aligns with common programming conventions. The next-protocol check enables the NF to apply protocol-specific processing logic, while the memory-bound check ensures that header accesses remain within packet bounds, which is necessary to pass the verifier. In practice, developers structure packet-parsing logic in this manner, and similar patterns can be observed in production systems such as Katran and Suricata.

Although this assumption holds for most NFs, we observe exceptions. In NF12 (\texttt{mptm\_decap\_xdp}), \oursystem\ is unable to identify the UDP protocol and its associated header fields. This is because the NF does not perform an explicit next-protocol check for UDP; instead, it treats all non-TCP traffic as UDP by default. 

When an NF is designed to operate on only a subset of traffic (e.g., TCP and UDP) and does not implement comprehensive protocol checks, \oursystem\ may fail to accurately infer the intended protocol semantics. This behavior highlights the limitation of the current version of our system. However this limitation could be addressed by extending \oursystem\ to handle NFs that are tailored to restricted traffic classes rather than fully general packet processing.

\myparab{\CFG generation time.} We evaluated the time taken by \oursystem\ to generate \CFG across various NFs, as shown in \fref{fig:time_rss_gen} (left) in~\secref{subsec:cfg_gc_fig}. The results are averaged over five runs and show a clear relationship between \CFG generation time and complexity of NFs. We observed that the time to generate a \CFG scales with the number of paths in the CFG. 
For NFs with $4$ to $4K$ paths, the generation time ranges from approximately $3$ to $300$ milliseconds.  
For NFs like NF13–NF16, which have the same number of paths but different numbers of instructions, the time proportionally increases with the increase in the number of instructions. We observe no significant difference in the time to extract network context for O3-optimized bytecode and other optimization flags.

% % ########################################
\begin{figure*}[t]
     \centering
     % \hspace{-0.7cm}
     \includegraphics [width=16cm, keepaspectratio]{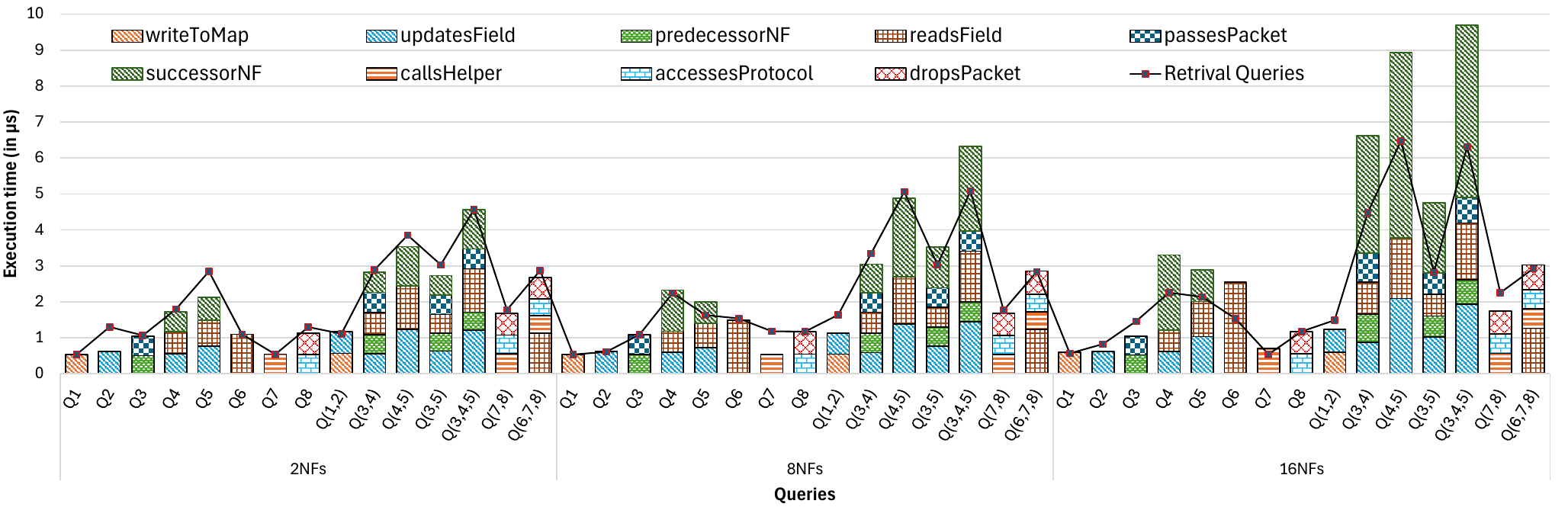}
     % \vspace{-0.6cm}
     \caption{Execution time: assertions (bars) and retrieval queries (line) on NF chains of varying length, where NFs are chosen randomly.}
     
     \label{fig:assertion_retrival_query_execution}
     % \Description{Execution time for the assertion (in bars) and retrieval queries (line) on NF chains of varying length.}
     % \vspace{-0.3cm}
 \end{figure*}

\subsection{Query execution}
\label{subsec:query_exec}
\label{subsec:eval_qe}

The Query Engine is built on the highly optimized \texttt{swi-prolog} runtime (V7.6.4)~\cite{swi_prolog}\footnote{\texttt{swi-prolog} implements several optimizations like indexing and unification to optimize the query resolution}; all of our query execution takes a few microseconds. We analyze the query execution time under different scenarios involving (i) assertion and retrieval, (ii) presence of recursive predicates, (iii) different numbers of \qpredslong (\secref{subsec:qe_eval}), and (iv) varying lengths of NF-chain (\secref{subsec:qe_eval}). To reduce the impact of noise, we run each query on the microbenchmarks $1000$ times, obtain a median for every $100$ runs, and then average the medians.

\myparab{Retrieval and Assertion queries.}
The execution time for each \qpredslong in the retrieval and assertion queries are shown in \fref{fig:assertion_retrival_query_execution}. We observe that the individual retrieval queries Q1(R)--Q8(R) and assertion queries Q1(A)--Q8(A) took $4\mu s$ when executed on the facts obtained from $16$ NFs. The combination variants of retrieval and assertion queries took $1.1\mu s$--$6.4\mu s$ and $1.1\mu s$--$9.6\mu s$, respectively.

\myparab{Recursive predicates.}
A rule is said to be recursive when the rule body includes a predicate that refers to the same rule head. For example, the predicates \texttt{successorNF} and \texttt{predecessorNF} shown in Table~\ref{tab:prolog_rules} are recursive predicates. We observe that the queries involving such recursive predicates take more time to execute than the non-recursive ones due to the additional overhead of handling recursive calls. 
These queries, when executed on the facts of sixteen NFs, took $1.04\mu s$--$3.3\mu s$ in comparison with the execution time of queries with no recursive predicates ($0.5\mu s$--$2.5\mu s$).

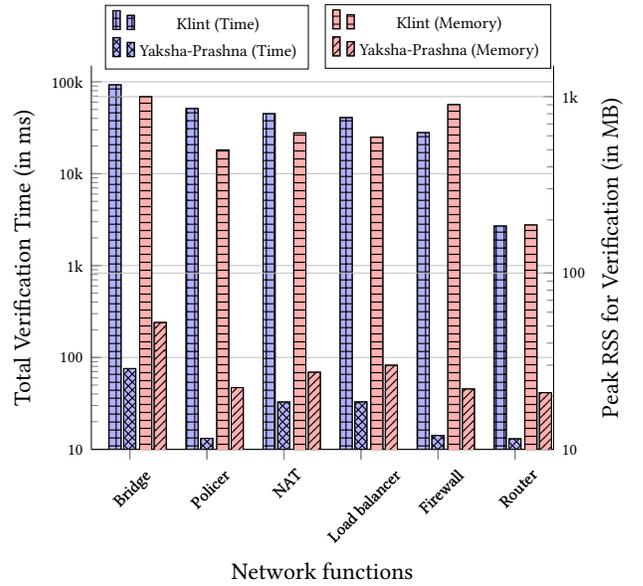
\begin{figure}[t]
 \centering
    \resizebox{\columnwidth}{!}{
        \begin{tikzpicture}
        
        \begin{axis}[
            % width=0.9\columnwidth,
            ybar,
            ymode=log,
            ytick={1e1,1e2,1e3,1e4,1e5},
            yticklabels={10,100,1k,10k,100k},
            ymin=10, ymax=150000,
            bar width=0.18cm,
            ylabel={Total Verification Time (in ms)},
            xlabel={Network functions},
            ymajorgrids=true,
            symbolic x coords={
                {Bridge },
                {Policer },
                {NAT },
                {Load balancer },
                {Firewall },
                {Router }
            },
            xtick=data,
            xtick pos=left,
            % x tick label style={font=\scriptsize, rotate=45, anchor=east},
            x tick label style={font=\scriptsize, rotate=45, xshift=2pt},
            y tick label style={font=\scriptsize},
            log ticks with fixed point,
            axis y line*=left,
            axis x line*=bottom,
            legend style={at={(0.25,0.99)},anchor=south,legend columns=1,font=\scriptsize},
            legend entries={Klint (Time),\archp (Time)},
            % bar shift auto
        ]
        % Klint times
        \addplot+[fill=blue!31, postaction={pattern=grid}, draw=black, bar shift=-0.22cm]
            coordinates {
                ({Bridge },92800)
                ({Policer },51200)
                ({NAT },45000)
                ({Load balancer },40800)
                ({Firewall },28000)
                ({Router },2696.2)
            };
        % Yaksha times
        \addplot+[fill=blue!31, postaction={pattern=crosshatch}, draw=black, bar shift=-0.002cm]
            coordinates {
                ({Bridge },75.69)
                ({Policer },13.19)
                ({NAT },32.77)
                ({Load balancer },32.85)
                ({Firewall },14.18)
                ({Router },13.026)
            };
        \end{axis}
		
		%%%%%%%%%%%%%%%%%%%%%%%%%%%%%%
		% Second dataset (right axis)
		\begin{axis}[
            % width=0.9\columnwidth,
            ybar,
            ymode=log,
            ytick={1e1,1e2,1e3},
            yticklabels={10,100,1k},
            ymin=10, ymax=1500,
            bar width=0.18cm,
            ylabel={Peak RSS for Verification (in MB)},
            ymajorgrids=true,
            symbolic x coords={
                {Bridge },
                {Policer },
                {NAT },
                {Load balancer },
                {Firewall },
                {Router }
            },
            xtick=data,
            xtick pos=left,
            % x tick label style={font=\scriptsize, rotate=45, anchor=east},
            x tick label style={font=\scriptsize, rotate=45, xshift=2pt},
            y tick label style={font=\scriptsize},
            log ticks with fixed point,
            axis y line*=right,
            axis x line*=none,
            legend style={at={(0.75,1.16)}, anchor=north,legend columns=1,font=\scriptsize},
            legend entries={Klint (Memory),\archp (Memory)}
            % bar shift=+1cm
        ]
        % Klint memory
        \addplot+[fill=red!31, postaction={pattern=horizontal lines}, draw=black, bar shift=+0.235cm]
            coordinates {
                ({Bridge },999.18)
                ({Policer },499.29)
                ({NAT }, 623.36)
                ({Load balancer },588.79)
                ({Firewall },901.42)
                ({Router },187.32)
            };
        % Yaksha memory
        \addplot+[fill=red!31, postaction={pattern=north east lines}, draw=black, bar shift=+0.45cm]
            coordinates {
                ({Bridge },52.41)
                ({Policer },22.39)
                ({NAT },27.39)
                ({Load balancer },30.01)
                ({Firewall },22.01)
                ({Router },20.96)
            };
        \end{axis}

    \end{tikzpicture}
    }
    \caption{Verification time (left) and Memory footprint (right).}
    \label{fig:Yaksha_klint_verification_time}
\vspace{-0.4cm}
\end{figure}

\myparab{Comparison with Klint.}
\fref{fig:Yaksha_klint_verification_time} (left) shows verification time of \archp and Klint on six representative NFs for 15 properties: P1–P15 (Appendix \secref{subsec:comp_with_klint_details}). \archp verification time includes CFG-NC generation and assertion execution, while for Klint it covers symbolic execution, invariant inference, and verification. Klint took 2.7 seconds to 1.5 minutes, whereas \archp finishes in $13$-$75$ ms, achieving a $200$-$1000$$\times$ speedup due to its lightweight dataflow analysis compared to Klint’s symbolic execution.

% % ################### NI Vs. No. of paths and NAI

% ##################

\section{Limitations}\label{limitations} 
In this section, we discuss the limitations of \archp. Section~\secref{subsec:discussion} in Appendix has more discussion points on identifying the packet-processing algorithms, complexity with random guesses for identifying network context information, and possible extensions to the language and the analyzer.

\myparab{Runtime queries.} NFs often determine packet actions based on header field values. For example, a load balancer uses flow information to forward subsequent packets to the same destination. 
The \archp does not support verification of such runtime properties, as they require runtime information, which is unavailable during the static analysis phase.

\myparab{Compliance assertions on map rules.} Maps are a key part of eBPF programs, but incorrect rules or missing actions can cause unexpected packet behavior. Operators often want to check the compliance of the map rules (e.g., absence of wildcard rules~\cite{p4v}), before insertion. Assuming map rules are given, \archp does not support these assertions. 

\myparab{Support for additional hookpoints.} Network functions can be attached to several hookpoints~\cite{eBPF_hooks} such as XDP, TC, and socket filters. Currently, \archp\ supports only XDP programs. Extending the tool to support additional hookpoints is primarily an engineering effort rather than a fundamental limitation of our approach, as differences across hooks (e.g., execution context structures, field offsets, and return codes) can be incorporated with additional implementation work.

%\vspace{-4mm}
\section{Related work}
\label{sec:related_work}

\myparab{\textit{Static analysis for verification.} }
Widely used eBPF verifiers like Linux eBPF verifier~\cite{verifier} and Prevail~\cite{prevail} use static analysis and abstract interpretation~\cite{cousot-cousot-1977-abstractinterpretation} to validate program termination, memory safety, and resource boundedness to ensure the kernel safety, but not the network context of the eBPF bytecode. Bromberger et al.~\cite{bromberger_verification_smt} propose verification of the functional properties of the eBPF program using Z3 SMT solver~\cite{z3-smt}.
Previous works have also used program verification for identifying bugs in P4~\cite{p4vera, datalog-verifying-p4}.  
Gravel~\cite{gravel_zhang2020automated}, Vigor~\cite{ vigor_zaostrovnykh2019verifying}, and VEP~\cite{vep_two_stage_verification} employ symbolic execution for verifying non eBPF-based NFs.  Beyond finding dependencies, \archp can be extended to verify network policies for stateful networks as in NetSMC~\cite{netsmc}. ASSERT-P4~\cite{p4assert} and DRACO~\cite{draco_lu2024towards} are Klee symbolic execution~\cite{cadar08klee} based approaches to verify P4/C programs.

\myparab{\textit{Static analysis for optimizations.} }
Merlin~\cite{merlin}, KFuse~\cite{kfuse}, K2~\cite{k2}, and Bonola et al.~\cite{program_warping_fpga_nics} propose static analysis based approaches for optimizing eBPF bytecode. Merlin performs eBPF aware optimizations on LLVM-IR, while KFuse and K2 design optimizations on the verified eBPF bytecode. Morpheus~\cite{morpheus} proposes runtime optimizations on eBPF programs based on the profiles of map access patterns.
On the other hand, \oursystem\ focuses on retrieving the network contexts of the eBPF bytecode.

\myparab{\textit{DSLs.} }
Domain-specific languages (DSLs) like Halide~\cite{halide-10.1145/2491956.2462176} (for image processing) and NetBlocks~\cite{NetBlocks-PLDI24} (for ad-hoc network protocol design) have helped create transpiler systems to obtain performance on the specific domains they target. \archp language is inspired by other DSLs like Datalog~\cite{datalog} and NDlog~\cite{ndlog} that are based on the declarative language paradigm.
These languages are designed to query and reason databases and networked systems in a distributed environment.
On the other hand, our language is designed to query and reason about the NF behavior of eBPF bytecodes.

\section{Conclusion}
We present \archp, a system for understanding eBPF-based NF bytecode and its interaction with other NFs. We propose (1) Analyzer to extract and model network context using dataflow analysis and propagation; and (2) Language, a simple but expressive domain-specific language for NF operators/developers to assert and retrieve eBPF bytecode behavior while abstracting out the complex low-level bytecode details. We evaluate (1) \archp language expressiveness by writing a wide range of queries checking NF properties, (2) \archp performance for $16$ XDP programs, and (3) compared \archp with state-of-the-art tools, Klint and DRACO.

\bibliographystyle{ACM-Reference-Format}
\bibliography{references}

\newpage
\appendix
\section{Appendix}

\subsection{\archp Assertion Execution}\label{subsec:AE}

In this section, we explain the workings of Analyzer and Query Engine using three assertions (\textbf{A1}, \textbf{A2} and \textbf{A3} defined in Sec.~\secref{subsec:motivating_examples}) executed on two bytecodes (Listing~\ref{lst:firewall_snippet} and~\ref{lst:update_mark} in Sec.~\secref{subsec:motivating_examples}).

Consider assertions \textbf{A1} and \textbf{A2} on the firewall bytecode (Listing~\ref{lst:firewall_snippet}). The analyzer builds bytecode's CFG and extracts network context by applying transfer function rules in~\tref{tab:inference_rules1}. More specifically, Rule \textbf{R4} extracts the fields updated by the bytecode, Rules \textbf{R11} and \textbf{R12} extract the list of protocols accessed, and Rule \textbf{R14} gathers the network context of all CFG nodes in each path. This information is stored in the knowledge base while maintaining the control flow structure in the bytecode. The Query Engine (QE) then executes the assertions: for \textbf{A1}, $updatesField$ retrieves the updated fields (\ie, TCP source port), causing the assertion to fail as the list is non-empty; for \textbf{A2}, $passes$ gathers the network context on all paths with XDP\_PASS as return action, and finally retrieves the list of protocols (\ie, TCP, UDP, and ICMP) processed in these paths. The assertion fails because ICMP is in the list. 

Next consider assertion \textbf{A3} on a chain of bytecodes, NF1 → NF2 → NF3, as shown in Listing~\ref{lst:update_mark}.  We assume the operator provides the order of NFs in a new or old chain. For instance, in the bpfman tool~\cite{bpfman}, NF priorities at a hook point define the order of their execution. Transfer function rules (\tref{tab:inference_rules1}) \textbf{R3} and \textbf{R4} extract the write list, that is, the list of fields updated by the bytecode, and the rules \textbf{R8} and \textbf{R9} extract the read list, the list of fields read by the bytecode. After applying transfer function rules to the bytecode, the extracted network context is stored in the knowledge base. While executing \textbf{A3} on the knowledge base, QE finds NF2's write list (\ie, \texttt{skb->mark} field) overlaps with the read list of NF3, which indicates dependency between NF2 and NF3. This understanding of the dependencies on the yet-to-be-deployed NF chain helps operators to find unexpected interactions and prevent outages before they occur. 

Using the bpfman tool~\cite{bpfman}, we deployed NF1 and NF3 at the XDP hook point, and executed assertion \textbf{A3} before deploying NF2. \fref{fig:bpfman-workflow} in Appendix~\ref{app:bpfman_deployment} illustrates the deployment of an example bytecode in the middle of an NF chain at a virtual interface.

\subsection{Header identification}\label{app:hid}
\begin{figure*}
     \centering
     % \hspace{-0.7cm}
     \includegraphics [width=\linewidth, keepaspectratio]{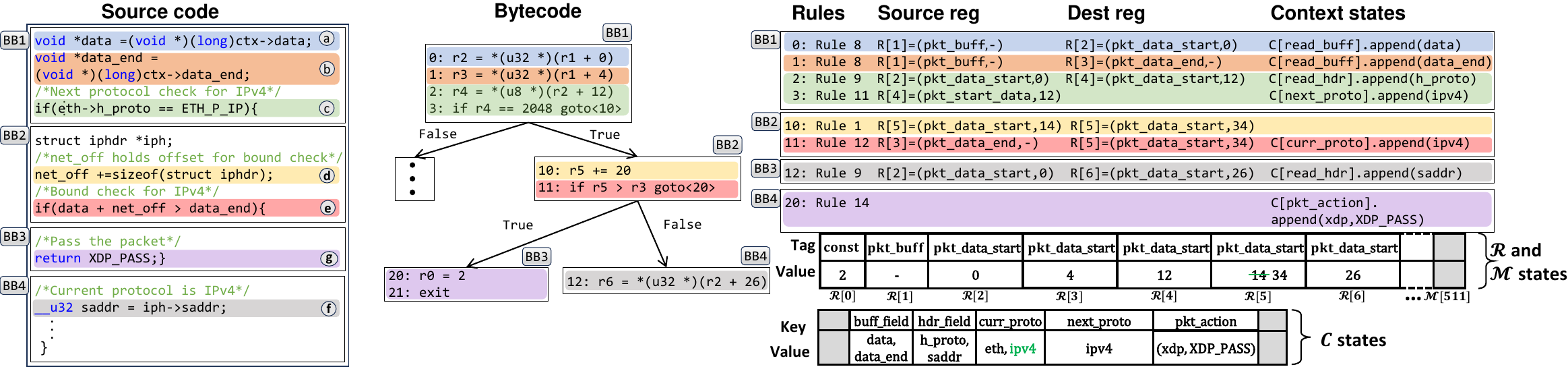}
     % \vspace{-0.6cm}
     \caption{Header identification using specific fields in the previously identified headers.}
   
     \label{fig:proto_running_example}
     % \Description{Header identification using specific fields in the previously identified headers.}
     \vspace{-0.3cm}
\end{figure*}
We apply rules R11 and R12 in~\tref{tab:inference_rules1} to find instructions that do header identification and memory-bound checks before accessing a header, respectively. More specifically, in R11, if the register points to a field\footnote{isTailField is a supporting function on spec that check whether the destination register refers to the current protocol's header field that identifies the next header.} that identifies the next header (\eg, ipv4.proto), then the next header's protocol is identified based on the immediate value (\eg, $6$ for TCP). A running example that illustrates these steps is shown in~\fref{fig:proto_running_example}. 
Here, we retrieve the next header's protocol using the `getProtoName\textsubscript{spec}' support function and store it in $\mathcal{C}$ with key as \customvalue{next\_proto}. Through R12, we confirm that the bytecode accesses the next header by ensuring that instructions associated with a memory-bound check are present. 
If present, we update the curr\_proto with the next\_proto. However, bytecode instructions that identify headers and perform memory-bound checks are spread across multiple basic blocks, because if a memory-bound check fails, the packet processing is halted. This requires forwarding protocol context (\ie, next\_proto and curr\_proto) to successor basic blocks, which we do through state propagation, as mentioned earlier.

\subsection{Discussion}\label{subsec:discussion}
\myparab{Runtime inputs.} \archp is designed to run queries on facts 
 representing the network context of eBPF bytecode. \archp's analyzer and language can be extended to understand or assert the packet-processing behavior at runtime for a specific packet header with a set of key-value entries in maps. For example, \textit{"All packets from a flow should be forwarded to the same destination"}. This helps to (1) debug connectivity and performance issues at runtime and (2) distinguish whether the unexpected packet-processing behavior (e.g., drop) is because of bugs in bytecode or an incorrect match rule in a map.

\myparab{Identifying packet-processing algorithms.} Currently, \archp analyzer extracts header or metadata fields accessed or updated by tracking packet buffer offsets. 
Due to the limitations of the static-analysis techniques~\cite{Rice53}, we are limited to scenarios that can be expressed as decidable rules.
It can be extended for identifying packet-processing algorithms (\eg, hash, DDoS detection, load balance, encryption) powered by ML techniques on the bytecode.
This enables NF audit that ensures the bytecode is processing packets as expected~\cite{auditbox_liu2021don} or provides opportunities to offload to accelerators~\cite{PPA}.

\myparab{Query engine.}  
 Due to the reliance on the \texttt{swi-prolog} runtime, \archp benefits from the decades of research on optimizations on query resolutions in 
 logical paradigm languages~\cite{phd/2009/wielemaker,robinson65logic, Horn1951-HOROSW}. 
 This results in a highly scalable and lightweight runtime system, where query execution barely takes a few microseconds with a small memory footprint.

\myparab{Comparison with random guess.} Identifying the network context becomes increasingly complex with random guesses due to the combinatorial space involved. Consider a bytecode with $p$ protocols with $f$ fields interacting with $m$ maps and $h$ helper functions. The user has to guess among $\binom{p_{max}}{p} \times \binom{f_{max}}{f} + \binom{m_{max}}{m} + \binom{h_{max}}{h}$ possible combinations. For instance, consider NF9 with $(p=6, f=10, h=2, m=9)$ and $(p_{max}=20, f_{max}=15, m_{max}=10, h_{max}=50)$, which amount to  $\binom{20}{6} \times \binom{15}{10} + \binom{10}{2} + \binom{50}{9} \approx 2.6\times10^{9}$ combinations. If the network context is unknown, the problem scales exponentially as the search space expands to $(2^{p_{max}-1}+1) \times (2^{f_{max}-1}+1) + (2^{m_{max}-1}+1) + (2^{h_{max}-1}+1) \approx 5.6\times10^{14}$  combinations (eBPF has $211$ helper functions, which amount to $2\times10^{15}$ and $1.64\times10^{63}$ combinations when $(p,f,m,h)$ are known and unknown, respectively). This emphasizes the challenges of identifying network context and the difficulty of random guessing, given the vast number of possibilities. By leveraging \archp, the time and effort can be significantly reduced, a big advantage over random guessing.

\subsection{\archp language predicates}
\label{app:predicates}
    \begin{enumerate}[leftmargin=4.5mm]
        \item \textbf{Field predicates} \textit{(field\_pred)}: Check if an NF reads or writes to a specific header or buffer field. The predicate returns the list of all fields read or updated if the field argument is a variable. 
        
        \item \textbf{Map predicates} \textit{(map\_operation\_pred, correlated\_map-\\\_pred):} The former predicate checks operations on maps, and the latter predicate checks for a possible correlation between the maps (\eg, map A lookup result is map B's key).
        
        \item \textbf{Action predicates} \textit{(pkt\_act\_pred):} Check if an NF performs actions like drop or redirect packets of a specific protocol. If protocol details are not mentioned, it returns the list of program paths ($P$) and the associated network context.  
        \item \textbf{Helper predicates} \textit{(helper\_pred):} Checks if an NF invokes a specific helper function. 
      
        \item \textbf{Protocol predicates} \textit{(protocol\_pred):} Checks if an NF processes a specific protocol. 
       
        \item \textbf{Order predicates} \textit{(order\_pred):} This predicate returns predecessor or successor NFs in an NF chain.
   
    \end{enumerate}
    
\subsection{Operational Semantics of \archp}
\label{app:opsemantcis}

\begin{table}[H]
\footnotesize
% \scriptsize
    \caption{Operational Semantics of the \archp.}
    \label{tab:prolog_rules}
    \resizebox{1\columnwidth}{!}{%
    \begin{tabular}{ll}
    \toprule
    \multicolumn{1}{c}{\textbf{Rule Head}} & \multicolumn{1}{c}{\textbf{Rule Body}} \\ \toprule
    % \multicolumn{1}{c}{\textbf{Rule}} \\
    % \midrule    
    \texttt{readsField(Nf, Fld):-} & \begin{tabular}[c]{@{}l@{}}\texttt{read\_header\_field(Nf, \_, Fld);}\\ \texttt{read\_buffer\_field(Nf, \_, Fld).} \end{tabular}\\ \midrule

    \texttt{updatesField(Nf, Fld):-} & \begin{tabular}[c]{@{}l@{}}\texttt{write\_header\_field(Nf, \_, Fld);}\\ \texttt{write\_buffer\_field(Nf, \_, Fld).} \end{tabular}\\ \midrule

    \texttt{successorNF(Nf, SNf):-} & \begin{tabular}[c]{@{}l@{}}\texttt{edge(Nf, SNf);edge(Nf, IntNf),}\\ \texttt{successorNF(IntNf, SNf).} \end{tabular}\\ \midrule

    \texttt{predecessorNF(Nf, PNf):-} & \begin{tabular}[c]{@{}l@{}}\texttt{edge(PNf, Nf); edge(PNf, IntNf),}\\ \texttt{predecessorNF(Nf, IntNf).} \end{tabular}\\ \midrule

    \texttt{passes(Nf, Hook, [(\texttt{Fld}, \texttt{Val})]):-} & \begin{tabular}[c]{@{}l@{}}\texttt{return\_action(Nf, Hook,}\texttt{"XDP\_PASS",}\\
    \texttt{[(Fld, val)]} \end{tabular}\\ \midrule

    \texttt{drops(Nf, Hook, [(\texttt{Fld}, \texttt{Val})]):-} & \begin{tabular}[c]{@{}l@{}}\texttt{return\_action(Nf, Hook,}\texttt{"XDP\_DROP",}\\
    \texttt{[(Fld, val)]} \end{tabular}\\ \midrule

    \begin{tabular}[c]{@{}l@{}}\texttt{mapWrite(Nf, Map,} \texttt{Fld):-}\end{tabular} & \texttt{write\_into\_map(Nf, \_, Fld, Map).} \\ \midrule

    \texttt{mapLookup(Nf, Map):-} & \texttt{read\_from\_map(Nf, \_, Map).} \\ \midrule

    \texttt{correlatedMaps(Nf, MapA, MapB):-} & \texttt{correlated\_map(Nf, \_, MapA, MapB).} \\ \midrule

    \begin{tabular}[c]{@{}l@{}}\texttt{accessesProtocol(Nf}, \texttt{ Fld, Val):-}\end{tabular} & \texttt{protocol\_accessed(Nf, \_, Fld, Val).} \\ \midrule

    \texttt{callsHelper(Nf, Helper):-} & \texttt{invoked\_helper(Nf, \_, Helper).} \\ \bottomrule
    \end{tabular}
    }
    \vspace{-0.25cm}
\end{table}

\subsection{Query execution workflow}\label{app:kbase}
We explain the steps in answering the queries using \\\texttt{updatesField(Nf, Fld)} query and associated facts in the knowledge base.

\begin{itemize}[leftmargin=10.5mm]
\item[\textbf{Rule:}] \texttt{updatesField(Nf, Fld):- write\_header\_field(Nf, \_, Fld)}.\par
\item[\textbf{Fact:}] \texttt{write\_header\_field("firewall\_kern\_kern/xdp", node\_3, ipv4.dst).}
\end{itemize}

     \myparab{\qpred matching:} 
    Each \qpredlong in a query is compared against the heads of the rules defined.
    The Query Engine evaluates the body of the rule corresponding to the match. All the predicates in the rule (\rpreds) must be satisfied for the rule to return true. 
    For the query above, the \qpred \texttt{updatesField} would be matched with the corresponding rule to evaluate the rule body.
    
   \myparab{Fact comparison:} The \rpreds are matched against the facts stored in the knowledge base. For a rule to be satisfied, each \rpredlong in the body must find a corresponding fact in the knowledge base. 
    In the above example,  the fact that the NF \texttt{firewall\_kern/xdp} updates the \texttt{ipv4.dst} field in the packet header would satisfy the rule for \texttt{updatesField}, abstracting the specific node \texttt{node\_3}. The underscore (`\_') in the rule serves as an anonymous variable, as the value of the node ID is not relevant for the match. This abstraction simplifies the query process by focusing only on whether the field \textit{Fld} is updated by the \textit{Nf}, regardless of the specific block where it is updated.

   \myparab{Handling variables:} If the query contains variables, the engine retrieves all values that satisfy the rule from the knowledge base (i.e., replacing the retrieved values would satisfy the \rpredlong of the rule's body). 
    For instance, if the above mentioned query is modified to ask which NFs update \texttt{ipv4.dst} (\ie, \texttt{updatesField(Nf,"ipv4.dst"}), the engine will return \texttt{firewall\_kern/xdp} as it satisfies the fact that it NF updates the destination IP field.
   
    \myparab{Result compilation and chaining:} After the matching process, results are compiled by collecting all the values that satisfy the query predicates. If a query involves multiple predicates, each of them is evaluated sequentially by following a logical AND/OR operation. 
    For example, if a query asks for NFs that both update a field (\texttt{updatesField}) and access a specific protocol (\texttt{accessesProtocol}), the engine first identifies NFs that update the field and then filters those based on protocol access, ensuring that only NFs meeting all criteria are returned.

\subsection{Retrieval and Assertion Queries}
\label{app:queries}
\begin{table}[H]
% \scriptsize
\footnotesize
\centering
\caption{List of Retrieval and Assertion Queries. The asterisk (*) denotes that the queries contain recursive predicates (\ie, successorNF and predecessorNF).}
\label{tab:queries-eval}
\resizebox{1\columnwidth}{!}{%
\begin{tabular}{llll} 
% p{0.125\linewidth}
\toprule
\multicolumn{2}{c}{\textbf{Retrieval Queries}} & \multicolumn{2}{c}{\textbf{Assertion Queries}} \\\toprule
Q1(R) & mapWrite(\textbf{Nf}, \textbf{\_} , \textbf{Fld}). & Q1(A) & mapWrite(\textbf{\textless{}NF3\textgreater{}}, \_ , \textbf{ipv4.src}). \\ \midrule
Q2(R) & updatesField(\textbf{Nf}, \textbf{Fld}). & Q2(A) & updatesField(\textbf{\textless{}NF8\textgreater{}}, \textbf{ipv4.dst}). \\ \midrule
Q3(R)* & \begin{tabular}[c]{@{}l@{}}predecessorNF(\textbf{Nf}, \textbf{SNf}), \\ passes(\textbf{SNf}, \textbf{Hook}, \textbf{[Fld, Arg]}).\end{tabular} & Q3(A)* & \begin{tabular}[c]{@{}l@{}}predecessorNF(\textbf{\textless{}NF15\textgreater{}},\textbf{\textless{}NF10\textgreater{}}),\\  passes(\textbf{\textless{}NF10\textgreater{}}, \textbf{xdp}, \textbf{[*,*]}).\end{tabular} \\\midrule
Q4(R)* & \begin{tabular}[c]{@{}l@{}}updatesField(\textbf{Nf},\\\textbf{xdp\_md.data}), \\successorNF(\textbf{Nf}, \textbf{SNf}), \\ readsField(\textbf{SNf},\textbf{xdp\_md.data}).\end{tabular} & Q4(A)* & \begin{tabular}[c]{@{}l@{}}updatesField(\textbf{\textless{}NF15\textgreater{}}, \textbf{ipv4.src}), \\ successorNF(\textbf{\textless{}NF15\textgreater{}}, \textbf{\textless{}NF12\textgreater{}}), \\ readsField(\textbf{\textless{}NF12\textgreater{}}, \textbf{ipv4.src}).\end{tabular} \\\midrule
Q5(R)* & \begin{tabular}[c]{@{}l@{}}updatesField(\textbf{Nf},\textbf{Fld}), \\ successorNF(\textbf{Nf},\textbf{SNf}), \\ readsField(\textbf{SNf},\textbf{Fld}).\end{tabular} & Q5(A)* & \begin{tabular}[c]{@{}l@{}}updatesField(\textbf{\textless{}NF10\textgreater{}}, \textbf{xdp\_md.data}), \\ successorNF(\textbf{\textless{}NF10\textgreater{}}, \textbf{\textless{}NF12\textgreater{}}), \\ readsField(\textbf{\textless{}NF12\textgreater{}},\textbf{xdp\_md.data}).\end{tabular} \\\midrule
Q6(R) & \begin{tabular}[c]{@{}l@{}}readsField(\textbf{Nf},\textbf{Fld}),\\readsField(\textbf{Nf},\textbf{Fld}).\end{tabular} & Q6(A) & \begin{tabular}[c]{@{}l@{}}readsField(\textbf{\textless{}NF6\textgreater{}},\textbf{tcp.flags.syn}), \\ readsField(\textbf{\textless{}NF6\textgreater{}},\textbf{tcp.flags.ack}).\end{tabular} \\\midrule
Q7(R) & callsHelper(\textbf{Nf}, \textbf{Helper}). & Q7(A) & \begin{tabular}[c]{@{}l@{}}callsHelper(\textbf{\textless{}NF3\textgreater{}},\\\textbf{bpf\_map\_update\_elem}).\end{tabular} \\\midrule
Q8(R) & \begin{tabular}[c]{@{}l@{}}accessesProtocol(\textbf{Nf},\textbf{Fld},\textbf{Proto}),\\drops(\textbf{Nf},\textbf{Hook},\textbf{[Fld,Arg]}).\end{tabular} & Q8(A) & \begin{tabular}[c]{@{}l@{}}accessesProtocol(\textbf{\textless{}NF15\textgreater{}},\textbf{eth.type},\textbf{ipv4}), \\ drops(\textbf{\textless{}NF15\textgreater{}}, \textbf{xdp},\textbf{[*,*]}).\end{tabular} \\\bottomrule
\end{tabular}%
}
\end{table}

\subsection{Bytecode deployment using bpfman}
\label{app:bpfman_deployment}

\begin{figure}[H]
    \centering
    \begin{subfigure}[t]{\linewidth}
        \centering
        \includegraphics[width=\linewidth]{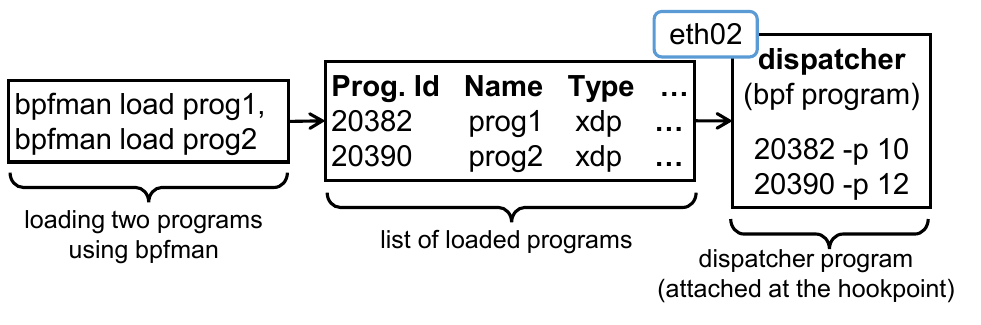}
        \caption{Two eBPF programs are attached with priorities (-p) 10 and 12.}
        \label{fig:bpfman-a}
    \end{subfigure}
    \hfill
    \begin{subfigure}[t]{\linewidth}
        \centering
        \includegraphics[width=\linewidth]{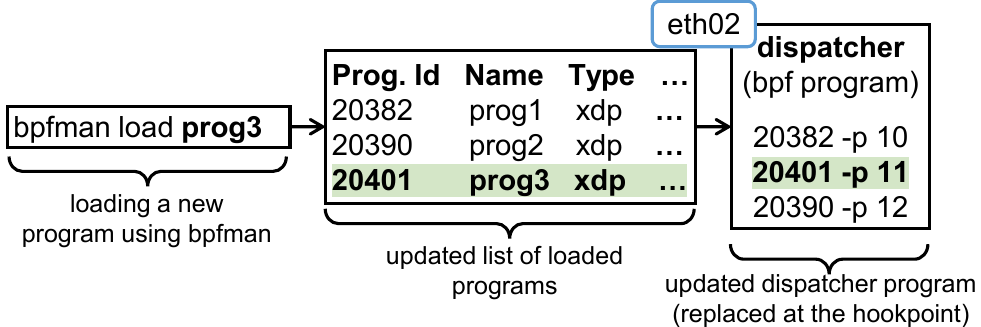}
        \caption{A new eBPF program is attached with priority (-p) 11.}
        \label{fig:bpfman-b}
    \end{subfigure}
    \caption{Workflow of bpfman tool for attaching multiple eBPF programs to a hook point (i.e., "eth02" interface).}
    \label{fig:bpfman-workflow}
    % \vspace{-0.3cm}
\end{figure}

Bpfman is a tool to manage eBPF programs on local hosts and Kubernetes. It supports load, unload, attach, and detach eBPF programs. It also supports the coexistence of multiple eBPF programs on a single interface with the help of the libxdp library~\cite{libxdp}. The workflow of bpfman is illustrated in the~\fref{fig:bpfman-workflow}. The eBPF bytecodes are loaded into the kernel and attached to a specific hook point. At a hook point, bpfman uses a special management program called a dispatcher to execute the programs in a specific order based on the priorities (a small number has high priority) set by the operator for each bytecode. The dispatcher program maintains a configuration structure to record the number of attached bytecodes, their priorities, chain call actions, and other metadata. 

The chain call action decides whether the packet should continue processing or the processing should be terminated. When a new bytecode is attached, a new dispatcher is created with the updated list of programs. 
The system atomically swaps the old dispatcher with the updated one, ensuring seamless updates and enforcing the correct execution order without downtime.

\subsection{CFG generation time and memory overheads}\label{subsec:cfg_gc_fig}
 % #########################################

\begin{figure}[t]
 \centering
    \resizebox{0.85\columnwidth}{!}{
        \begin{tikzpicture}
        
        % First plot for the left y-axis
        \begin{axis}[
            ybar, 
            ymin=0, ymax=300,
            bar width=0.15cm, 
            ytick distance=50,
            ylabel={Time to generate CFG-NC (in ms)}, 
            xlabel={Network function}, 
            ymajorgrids=true, yminorgrids=true, 
            symbolic x coords={NF1,NF2,NF3,NF4,NF5,NF6,NF7,NF8,NF9,NF10,NF11,NF12,NF13, NF14, NF15, NF16},
            xtick=data, 
            xtick pos=left,
            x tick label style={font=\scriptsize, rotate=45}, 
            y tick label style={font=\scriptsize}, 
            axis y line*=left, 
            axis x line*=bottom, 
            bar shift=-0.1cm,
            legend style={at={(0.2,0.83)},anchor=south,legend columns=-1, , font=\footnotesize}, 
            legend entries={Time},
        ]
        % First dataset (left axis)
        \addplot+[fill=red!31, postaction={pattern=grid}, draw=none, % Remove the borders from bars
   error bars/.cd,
    y dir=both, y explicit, error bar style={line width=2.5pt, color=black},
] plot coordinates {
    (NF1,3.64) +-    (0.07,0.07) 
    (NF2,4.41) +-    (0.05,0.05) 
    (NF3,4.67) +-    (0.01,0.01) 
    (NF4,9.01) +-    (0.14,0.14) 
    (NF5,9.41) +-    (0.07,0.07) 
    (NF6,9.78) +-    (0.18,0.18) 
    (NF7,11.28) +-   (0.49,0.49) 
    (NF8,14.42) +-   (0.28,0.28) 
    (NF9,54.38) +-   (0.88,0.88) 
    (NF10,56.06) +-  (1.06,1.06)
    (NF11,93.49) +-  (0.64,0.64) 
    (NF12,133.65) +- (1.29,1.29) 
    (NF13,244.00) +- (1.10,1.10) 
    (NF14,254.92) +- (2.85,2.85) 
    (NF15,255.83) +- (0.80,0.80) 
    (NF16,269.38) +- (1.73,1.73)
        };
        \end{axis}
        
        % Second plot for the right y-axis
        \begin{axis}[
            ybar, 
            ymin=0, ymax=65,
            bar width=0.15cm, 
            ytick distance=10, 
            ylabel={Peak RSS for generating CFG-NC (MB)}, 
            ymajorgrids=true, yminorgrids=false, 
            axis y line*=right, 
            axis x line=none, 
            symbolic x coords={NF1,NF2,NF3,NF4,NF5,NF6,NF7,NF8,NF9,NF10,NF11,NF12,NF13, NF14, NF15, NF16}, 
            y tick label style={font=\scriptsize}, 
            nodes near coords, 
            every node near coord/.append style={rotate=90, anchor=west, font=\scriptsize}, 
            bar shift=0.1cm,
            legend style={at={(0.5,0.83)},anchor=south,legend columns=-1, font=\footnotesize}, 
            legend entries={Peak RSS},
        ]
        % Second dataset (right axis)
        \addplot[fill=blue!31, postaction={pattern=crosshatch}, draw=none, point meta=explicit symbolic,] coordinates {
        (NF1,5.18)
        (NF2,5.22)
        (NF3,5.22)
        (NF4,5.48)
        (NF5,5.58)
        (NF6,5.78)
        (NF7,5.68)
        (NF8,5.99)
        (NF9,11.13)
        (NF10,11.40)
        (NF11,19.54)
        (NF12,24.37)
        (NF13,40.36)
        (NF14,43.17)
        (NF15,47.09)
        (NF16,46.39)
        };
        \end{axis}
        
    \end{tikzpicture}
    }
    \caption{Time (left y-axis) and Memory (right y-axis) consumed for generating \CFG \vspace{-4mm}}
    
    \label{fig:time_rss_gen}
    % \Description{A bar plot showing Time (in left y-axis) and Memory (in right y-axis) consumed for generating \CFG}
% \vspace{-0.4cm}
\end{figure}
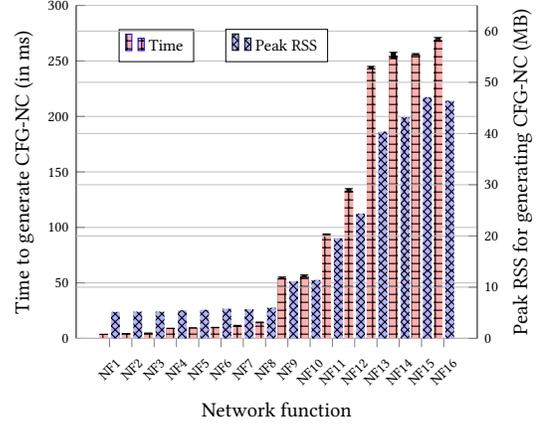
% #########################################

\subsection{Comparison Overview}
\label{subsec:comp_overview}
We compared \archp with two state-of-the-art
works: Klint~\cite{verify_nf_binaries} and DRACO~\cite{draco_lu2024towards}. Klint is a verification tool that
checks the conformance of NF binaries with the specification
written in Python. NF can be written in different languages,
such as C or Rust, using frameworks such as DPDK or eBPF.
DRACO is another verification tool that verifies eBPF-based NFs using specifications and assertions, but requires the availability of source code to do so. The comparison is organized into two parts: one dedicated to Klint and the other to DRACO, each addressing the corresponding class of NFs and properties used for verification.

% ###############################################

\begin{table*}[]
\centering
\caption{Assertions for the properties and Retrieval Queries written using \archp Language.}
\label{tab:literature-queries-new}
\footnotesize
\begin{tabular}{|l|l|l|}
\hline
\textbf{S.No.} & \begin{tabular}[c]{@{}l@{}}\textbf{Properties (P1-P22)/Retrieval Qs (Q23-24)}\end{tabular}                                                                              & \textbf{\archp assertions/Retrieval Qs}                                                                                                                                                                                                                                                                                                                                             \\ \hline
P1              &  Expected header order                                                                                                                        & \begin{tabular}[c]{@{}l@{}}\texttt{!passes("nf\_id", xdp, {[}(var, var){]}), accessesProtocol("nf\_id",\_, "eth"),}\\ \texttt{accessesProtocol("nf\_id",}\texttt{"eth.type","ipv4")}.\end{tabular}                                                                                                                                                                                                                \\ \hline
P2              &  Validity of IPv4 header                                                                                                                      & \begin{tabular}[c]{@{}l@{}} \texttt{!passes("nf\_id", xdp, {[}("ipv4.ihl", >= 5),("ipv4.ihl", <= "ipv4.totlen"),} (...)\texttt{{]}).}\end{tabular}                                                                                                                                                                                                                                                                                                                                 \\ \hline
P3              & \begin{tabular}[c]{@{}l@{}} Expected header field updates by NF\end{tabular}                                                               & \begin{tabular}[c]{@{}l@{}} \texttt{!passes("nf\_id", xdp, {[}(var, var){]}),} \texttt{"ipv4.ttl").} \end{tabular}                                                                                                                                                                                                                                                                     \\ \hline
P4              &  Longest prefix matching                                                                                                                      & \begin{tabular}[c]{@{}l@{}} \texttt{!passes("nf\_id", xdp, {[}(var, var){]}),} \texttt{mapLookup("nf\_id", "router\_map").} \end{tabular}                                                                                                                                                                                                                                                                       \\ \hline
P5              & \begin{tabular}[c]{@{}l@{}} Packet content preservation\end{tabular}                                                                       & \texttt{!updatesField("nf\_id",  *).}                                                                                                                                                                                                                                                                                                                                                              \\ \hline
P6              &  Filtering external flows                                                                                                                     & \begin{tabular}[c]{@{}l@{}} \texttt{!passes("nf\_id",xdp,{[}("xdp\_md.ingress\_ifindex", EXTERNAL), (var, var){]}),} \\\texttt{mapLookup("nf\_id",}\texttt{"firewall\_map").} \end{tabular}                                                                                                                                                                                                                        \\ \hline
P7              & \begin{tabular}[c]{@{}l@{}} Remembering internal flows\end{tabular}                                                                        & \begin{tabular}[c]{@{}l@{}} \texttt{!passes("nf\_id", xdp, {[}("xdp\_md.ingress\_ifindex",} \texttt{!EXTERNAL),(var, var){]}),}\\\texttt{mapWrite("nf\_id", "firewall\_map", *).} \end{tabular}                                                                                                                                                                                                                     \\ \hline
P8              & \begin{tabular}[c]{@{}l@{}} Update the source IP and port of the internal \\flow\end{tabular}                                                & \begin{tabular}[c]{@{}l@{}}  \texttt{!passes("nf\_id", xdp, {[}("xdp\_md.ingress\_ifindex", !EXTERNAL), (var, var){]}),} \\ \texttt{updatesField("nf\_id",}\texttt{ipv4.src), updatesField("nf\_id", "tcp.sport"); } \\\texttt{updatesField("nf\_id", "udp.sport").} \end{tabular}                                                                                                                                            \\ \hline
P9              & \begin{tabular}[c]{@{}l@{}} Restore the destination IP and port of the\\ external flow\end{tabular}                                       & \begin{tabular}[c]{@{}l@{}} \texttt{!passes("nf\_id", xdp, {[}("xdp\_md.ingress\_ifindex", EXTERNAL),(var, var){]}), } \\ \texttt{updatesField("nf\_id", "ipv4.dst"), } \texttt{updatesField("nf\_id", "tcp.dport");}\\ \texttt{updatesField("nf\_id", "udp.dport").} \end{tabular}                                                                                                                                             \\ \hline
P10             & \begin{tabular}[c]{@{}l@{}}Selected transmission port must not be the\\ingress port\end{tabular}                                           & \begin{tabular}[c]{@{}l@{}} \texttt{!passes("nf\_id", xdp, {[}(var,var){]}), mapLookup("nf\_id", "bridge\_map"),} \\ \texttt{readsField("nf\_id", "xdp\_md.ingress\_ifindex").} \end{tabular}                                                                                                                                                                                                               \\ \hline
P11             & Broadcast if unknown                                                                                                                         & \begin{tabular}[c]{@{}l@{}} \texttt{!drops("nf\_id", xdp, {[}(var, var){]}),} \texttt{mapLookup("nf\_id", "bridge\_map").} \end{tabular}                                                                                                                                                                                                                                                                                                                    \\ \hline
P12             & The learning process                                                                                                                         & \begin{tabular}[c]{@{}l@{}} \texttt{!passes("nf\_id", xdp, {[}(var, var){]}),} \texttt{mapWrite("nf\_id", “bridge\_map”, *).} \end{tabular}                                                                                                                                                                                                                                                                   \\ \hline
P13             & External traffic policing                                                                                                                    & \begin{tabular}[c]{@{}l@{}} \texttt{!drops("nf\_id", xdp, {[}("xdp\_md.ingress\_ifindex",} \texttt{!EXTERNAL){]}).} \end{tabular}                                                                                                                                                                                                                                                                                              \\ \hline
P14             & Backend liveness                                                                                                                             & \begin{tabular}[c]{@{}l@{}} \texttt{!drops("nf\_id", xdp,{[}("xdp\_md.ingress\_ifindex", !EXTERNAL), (var, var){]}),} \\ \texttt{callsHelper("nf\_id", } \texttt{"bpf\_ktime\_get\_ns"), mapWrite("nf\_id",}\texttt{"backend\_map",*).} \end{tabular}                                                                                                                                                                                         \\ \hline
P15             & \begin{tabular}[c]{@{}l@{}}Load balancing external traffic\end{tabular}                                                                   & \begin{tabular}[c]{@{}l@{}} \texttt{!passes("nf\_id", xdp, {[}("xdp\_md.ingress\_ifindex", EXTERNAL), (var, var){]}),} \\ \texttt{mapLookup("nf\_id", "backends\_map").}\end{tabular}                                                                                                                                                                                                                                      \\ \hline
P16             & \begin{tabular}[c]{@{}l@{}}Drop all fragmented packets\end{tabular}                                                                       & \begin{tabular}[c]{@{}l@{}} \texttt{passes("nf\_id", xdp, {[}(var, var){]}),} \texttt{readsField("nf\_id", "ipv4.frag").} \end{tabular}                                                                                                                                                                                                                                                                   \\ \hline
P17             & \begin{tabular}[c]{@{}l@{}}Forward all ICMP echo request packets\end{tabular}                                                             & \begin{tabular}[c]{@{}l@{}} \texttt{passes("nf\_id", xdp, {[}(var, var){]}),} \texttt{readsField("nf\_id", "icmp.request"),} \\ \texttt{updateField("nf\_id", "icmp.*"), updatesField("nf\_id", "ipv4.*"), } \\ \texttt{updatesField("nf\_id", "eth.*").} \end{tabular}                                                                                                                                                         \\ \hline
P18             & \begin{tabular}[c]{@{}l@{}}Handle only TCP SYN packets\end{tabular}                                                                       & \begin{tabular}[c]{@{}l@{}} \texttt{passes("nf\_id", xdp, {[}(var, var){]}),} \texttt{readsField("nf\_id", "tcp.syn").} \end{tabular}                                                                                                                                                                                                                                                                          \\ \hline
P19             & \begin{tabular}[c]{@{}l@{}}Add custom redirection header\end{tabular}                                                                     & \begin{tabular}[c]{@{}l@{}} \texttt{passes("nf\_id",   xdp, {[}(var,var){]}),callsHelper("nf\_id","bpf\_xdp\_adjust\_head"),} \\ \texttt{updatesField("nf\_id","tcp.dataoffset").} \end{tabular}                                                                                                                                                                                                               \\ \hline
P20             & \begin{tabular}[c]{@{}l@{}}Forward all packets unchanged\end{tabular}                                                                     & \begin{tabular}[c]{@{}l@{}} \texttt{passes("nf\_id",   xdp, {[}(*, *){]}). AND !updatesField("nf\_id", *).} \end{tabular}                                                                                                                                                                                                                                                                           \\ \hline
P21             & Correlated maps                                                                                                                              & \texttt{correlatedmaps({[}"map\_one", "map\_two"{]}).}                                                                                                                                                                                                                                                                                                                                              \\ \hline
P22             & \begin{tabular}[c]{@{}l@{}}NF dependency: Read after write (RAW),\\ Write after read (WAR) and Write after\\ write (WAW)\end{tabular}   & 

\begin{tabular}[c]{@{}l@{}}RAW assertion: \texttt{updatesField("nf\_id1", *), successorNF("nf\_id1", "nf\_id2"),} \\ \texttt{readsField("nf\_id2", *).} \\ WAR assertion:\texttt{ readsField("nf\_id1", *), successorNF("nf\_id1", "nf\_id2"),} \\ \texttt{updatesField("nf\_id2", *).} \\ WAW assertion: \texttt{updatesField("nf\_id1", *), successorNF("nf\_id1", "nf\_id2"),} \\ \texttt{updatesField("nf\_id2", *).} \end{tabular} \\ \hline
Q23             & \begin{tabular}[c]{@{}l@{}}Find the list of packet fields updated by NF\end{tabular}                                                  & \begin{tabular}[c]{@{}l@{}} \texttt{updatesField("nf\_id", Fld).} \end{tabular}                                                                                                                                                                                                                                                                                          \\ \hline
Q24             & \begin{tabular}[c]{@{}l@{}}Find the list of packet fields where one NF\\ updates and its successor reads \\(RAW dependent)\end{tabular} & \begin{tabular}[c]{@{}l@{}} \texttt{updatesField(“nf\_id1”, Fld), successorNF(“nf\_id1”, “nf\_id2”),} \\ \texttt{readsField(“nf\_id2”, Fld).} \end{tabular}                                                                                                                                                                                                                                                 \\ \hline
\end{tabular}
\end{table*}

% ################################################

\subsubsection{Details of implemented NFs and properties extracted for comparison with Klint}
\label{subsec:comp_with_klint_details}
For comparison, we evaluated NF’s conformance to their specifications, similar to Klint’s verification of C-based binaries. For this comparison, \archp needed (1) eBPF bytecode of the NFs and (2) the properties used for the verification. Since Klint’s NFs are written in C, we wrote semantically equivalent eBPF-based NFs and used their bytecode for comparison. These are six representative NFs: Router, NAT, Firewall, Bridge, Load Balancer, and Policer. We validated semantic equivalence between the original and rewritten NFs; the validation methodology is described in Appendix~\ref{subsec:semantic_equivalence_of_rewritten_eBPF_programs}. The properties to verify were extracted from the Klint specification. All properties fall into Category 1 (\secref{subsec:motivating_examples}). Detailed description of the developed NFs and the extracted properties is discussed below. The \archp assertions used to assert these properties can be found in \tref{tab:literature-queries-new}.

\myparab{(1) Router:} The developed router is a basic IPv4 router that forwards the valid IP packets based on the routing table. The routing table is implemented using an eBPF map that maps the destination IP addresses to the interface.
The router was verified using the following key properties:-

\myparab{Property 1: Expected header order.} An IPv4 router should only allow Ethernet followed by IPv4 packets. All other packets (e.g. IPv6 packets) must be dropped. This property helps to take protocol-specific packet actions. To verify this property, we asserted a high-level assertion,\textit{“Only Ethernet and IPv4 packets should be forwarded”}, whose exact assertion is presented in row P1 of \tref{tab:literature-queries-new}.

\myparab{Property 2: Validity of IPv4 header.} 
A router must ensure the validity of an IPv4 header before processing the packet. This is achieved by verifying properties such as: (1) the IP header length is at least 20 bytes, (2) the IP header length does not exceed the total length of the packet, (3) the checksum is correct, and (4) the TTL is non-zero, etc. These validity checks are crucial for identifying and discarding malformed IP packets. We used high-level assertions such as \textit{"Every outgoing packet must have an IP header length of at least 20 bytes"}, \textit{"Every outgoing packet must have an IP header length less than total length"}, etc, to verify these properties. The exact assertion is presented in row P2 of \tref{tab:literature-queries-new}.

\myparab{Property 3: Expected header field updates by NF.} One of the key properties of a router is to decrement the time-to-live (TTL) field of every outgoing packet. This property ensures that the packet does not circulate indefinitely within the network.
We verified this property using the high-level assertion, \textit{“IPv4 ttl must be updated on every outgoing packet”}, whose exact assertion is presented is row P3 of \tref{tab:literature-queries-new}.

\myparab{Property 4: Longest-prefix-match.} An IPv4 router must perform longest-prefix-match (LPM) while mapping the destination IP address to the appropriate forwarding interface. LPM enables faster and more efficient matching than alternative approaches such as exact matching. eBPF provides a specific map type to perform LPM matching. To verify this property, we wrote two high-level assertion, \textit{“All the outgoing packets must perform a lookup on router map"}, and \textit{"The map type of the router map must be LPM”}. Since the \archp language lacks the predicate to assert for the map types, we could only use the first assertion for verification. The exact assertion is presented in row P4 of \tref{tab:literature-queries-new}. 

\myparab{(2) Firewall:} The developed firewall implements stateful traffic filtering, where the state is captured using a map. It primarily performs two tasks: (1) blocking unknown external flows, and (2) remembering internal flows, when necessary, and forwarding them. To verify the correctness of the firewall, we used the following three key properties:-

\myparab{Property 5: Packet content preservation.} Typically, a firewall should not modify the contents of the packets it forwards. This property is important for ensuring the integrity of the forwarded packets. We used the assertion A1, discussed in the motivation section (\secref{subsec:motivating_examples}), to verify this property, which is also presented in row P5 of~\tref{tab:literature-queries-new}.

\myparab{Property 6: Filtering external flows.} One of the key properties of a firewall is to block unknown external packets. This property is crucial as it ensures the safety of the internal network. To verify the same, we asserted, \textit{“All external packets must not be forwarded, if no entry is found in the firewall map”}. The exact assertion is presented in row P6 of~\tref{tab:literature-queries-new}.

\myparab{Property 7: Remembering internal flows.} For a stateful firewall to function properly, it must remember the state of the internal flows. This property is also crucial for allowing the “remembered” external flows (Property 6) in the network. To assert this property, we used a high-level assertion, \textit{“For all internal packets, if the flow entry is not present in the firewall map, the map must be updated with flow information”.} The exact assertion is presented in row P7 of~\tref{tab:literature-queries-new}.

\myparab{(3) NAT:} The developed stateful NAT performs two primary functions: (1) for internal flows, it updates the source IP address and port with the public IP address and the masked port number, and (2) for external flows, it restores the destination IP address and port to the original internal IP address and port number. We also verified \textbf{Property 6} and \textbf{Property 7} on NAT, since they are essential for ensuring the safety of the internal network and guaranteeing the correct stateful behavior of the NAT. Apart from that, two more key properties were verified:-

\myparab{Property 8: Update the source IP and port of the internal flow.} For any internal flow, the NAT must update the source IP address to the public IP and mask the source port. This property is important because it allows multiple internal devices to be represented by a single public IP address, while still keeping their flows uniquely identifiable with the help of the masked port number. To verify the same, we used a high-level assertion, \textit{“Source IP address and port of all the internal packets must be updated”}, whose exact assertion is presented in row P8 of~\tref{tab:literature-queries-new}.

\myparab{Property 9: Restore the destination IP and port of the external flow.} For any external flow, the NAT must restore the destination IP and port to the original internal private address and port. This is important to ensure that the external packets reaching NAT can be directed to the correct internal device. To verify the same, we used a high-level assertion, \textit{“Destination IP address and port of all the known external packets must be updated”}. The exact assertion is presented in row P9 of~\tref{tab:literature-queries-new}. 

\myparab{(4) Bridge:} The developed bridge is a MAC-learning bridge that forwards the Ethernet frame based on the destination MAC address. The mapping of the destination MAC and interface is stored using a map. The bridge was verified using the following three key properties:-

\myparab{Property 10: Selected transmission port must not be the ingress port.} For a bridge, the selected transmission port must not be the same as the ingress port of the received frame. This check is essential to prevent unnecessary loops in the network. We verified this property using the high-level assertion \textit{"The transmission port of the outgoing packet must not be equal to the ingress port"}. The exact assertion is presented in row P10 of~\tref{tab:literature-queries-new}.  

\myparab{Property 11: Broadcast if unknown frame.} If the frame received by the bridge does not match any entry in the forwarding table, then the frame must not be dropped and should be broadcast to all the interfaces, except the ingress interface. This property guarantees that the frame can still reach its intended destination, even when the bridge lacks knowledge of the appropriate outgoing interface. To verify this, we used two high-level assertions: \textit{“If map lookup fails on the bridge map, then the packet must not be dropped”}, and \textit{"The packet must be broadcast, except for the ingress"}. Since the \archp Language does not support predicates to assert for packet broadcast, we could only verify the first assertion. The exact assertion is presented in row P11 of \tref{tab:literature-queries-new}.

\myparab{Property 12: The learning process.} The learning process observes the source addresses of frames received on each port and updates the forwarding table. This process is important as it allows the bridge to build the forwarding table dynamically and forward frames efficiently, rather than relying on broadcasting. We verified this property using the high-level assertion, \textit{“If the source MAC of the received frame is not present in the bridge map, then the map must be updated accordingly”}. The exact assertion is presented in row P12 of~\tref{tab:literature-queries-new}.  

\myparab{(5) Policer:} The developed policer is a stateful network function that implements a per-destination token bucket for external packets, effectively limiting traffic based on the packet size. A map is used to keep track of the state, using the IP address as the key and storing the token bucket details, such as token size, as the associated value. We verified the following property on the policer:-

\myparab{Property 13: External traffic policing.} The policer should apply rate limiting only to external traffic, updating the state according to the packet size, while all internal packets must be forwarded. To verify the former property, we used the high-level assertion \textit{“All external packets must perform lookup on the policer map and update it accordingly”}, while the latter property was verified using \textit{“All internal packets must be forwarded”} assertion. The exact assertion is presented in row P13 of~\tref{tab:literature-queries-new}.

\myparab{(6) Load balancer:} The developed load balancer is an eBPF implementation of Google’s load balancer called “Maglev”. It picks one of the backends from the backend pool based on consistent hashing, and remembers the flow and backend mapping using a map. It also uses another map to keep track of the live backends, which is updated after receiving \textit{heartbeat} packets from the backends. We verified the load balancer with \textbf{Property 5}, as the load balancer should not modify the packet contents while processing the packet. Apart from that, we also verified the following two key properties:-

\myparab{Property 14: Backend liveness.} If the packet comes from backends, it must update the live backend map with the current time and then get dropped. This is important to keep track of live backends and to avoid forwarding packets to dead backends. We verified the property using the high-level assertion, \textit{“All the dropped backend packets must update the live backend map with the current time”}. The exact assertion is presented in row P14 of~\tref{tab:literature-queries-new}. 

\myparab{Property 15: Load balancing external traffic.} The external packets should be forwarded to one of the chosen backends, and if the backend is dead, the packet must be dropped. We verified the property using the high-level assertion, \textit{"All external packets must be forwarded to the backend if the backend is alive"}, whose exact assertion is presented in row 15 of \tref{tab:literature-queries-new}.

\subsubsection{Details of NFs and properties extracted for comparison with DRACO}
\label{subsec:comp_with_draco_details}

We compared Yaksha-Prashna with DRACO in terms of expressiveness. We extracted the properties used by DRACO for verifying the three real-world eBPF programs, Katran~\cite{katran}, CRAB~\cite{Crab_loadbalancer}, Fluvia~\cite{fluvia}, and one eBPF program from academic literature, hXDP FW~\cite{hXDP}. All the properties fall under Category 1 (\secref{subsec:motivating_examples}). 
A brief description of the NFs and the extracted properties is discussed below. The Yaksha-Prashna assertion can be found in \tref{tab:literature-queries-new}.

\myparab{Katran:} Katran is a Layer-4 load balancer developed by Meta that uses XDP for efficient packet processing. It is designed to provide scalable and low-latency connection balancing. Katran was verified using the following two key properties:-

\myparab{Property 16: Drop all fragmented packets}. One of Katran's constraints is its inability to handle fragmented packets; as a result, it drops all such packets. We expressed this property using a high-level assertion,\textit{"None of the fragmented packets should be forwarded"}, whose exact assertion is presented in row P16 of \tref{tab:literature-queries-new}.

\myparab{Property 17: Forward all ICMP echo request packets.} Katran takes many protocol-specific packet actions. One such action handles ICMP echo request packets. When encountering such request packets, Katran converts them into echo replies by updating the relevant ICMP, IP and Ethernet header fields, and forwards them. We expressed it using \textit{“All the ICMP echo request packets must be forwarded after updating relevant fields of ICMP, IP and Ethernet header”} high-level assertion. The exact assertion is presented in row P17 of \tref{tab:literature-queries-new}.

\myparab{CRAB:} CRAB is another Layer-4 load balancer that participates only during connection setup. It handles SYN packets to choose a backend server and lets clients and servers talk directly via \textit{Connection Redirection}, thus lowering latency and avoiding bottlenecks while allowing complex stateful load balancing. We verified CRAB using the following two key properties:-

\myparab{Property 18: Handle only TCP SYN packets.} CRAB participates only in the connection establishment phase, where it handles TCP SYN packets and discards all the other packets. To express this property, we used the high-level assertion,\textit{“All forwarded packets must be TCP SYN packets”}, whose exact assertion is presented in row P18 of \tref{tab:literature-queries-new}.

\myparab{Property 19: Add custom redirection header.} To stay off the datapath between the client and the server, CRAB leverages \textit{Connection Redirection}, where the communication between client and server happens directly without involving the load balancer. A custom redirection header is appended to all the outgoing packets to enable this feature. We expressed the property using the high-level assertion, “All the outgoing packets must have a custom header appended”. The exact assertion is presented in row P19 of \tref{tab:literature-queries-new}.

\myparab{Fluvia:} Fluvia is an IPFIX exporter that leverages XDP to collect metadata related to the incoming IPv6 flows, and exports it to the userspace for analysis. Fluvia was verified using the following property:-

\myparab{Property 20: Forward all the packets unchanged.} A flow exporter should transparently collect the flow's metadata without dropping any incoming packets. Moreover, it should also not modify the packet contents. We used the high-level assertion, \textit{"All packets must be passed"}, in conjunction with \textbf{Property 5} to express the same. The exact assertion is presented in row P20 of \tref{tab:literature-queries-new}.

\myparab{hXDP Firewall:} hXDP firewall is a simple stateful firewall from the academic literature\cite{hXDP}, which implements similar functionality as the firewall discussed in Section~\secref{subsec:comp_with_klint_details}. Apart from verifying the key firewall properties (Property 5-7), we also verified the following specific property, which is specific to the hXDP firewall implementation:-

\myparab{Property 21: Correlated Maps.} Correlated maps are often used to share state across different maps. Two maps are considered correlated when the return value of one map operation is used as the argument for another map operation, such as a lookup or update. Lines 13-15 of Listing \ref{lst:firewall_snippet} capture an instance of correlated maps, where the value from map lookup on \textit{flow\_ctx\_table} is stored in \textit{flow\_leaf} (line 13), and later used as a key for redirecting the packet using the \textit{tx\_port} map (line 15). We have a dedicated map predicate \textit{correlated\_maps} to capture the correlated maps, which is presented in row P21 of \tref{tab:literature-queries-new}.

In addition to the aforementioned properties, DRACO also asserted a few generic network function properties, such as “Ethernet source address remains unchanged across all packets”, “All IPv6 packets should be dropped”, etc. These properties could likewise be expressed using the \archp Language.

\subsection{Query execution}\label{subsec:qe_eval}
\myparab{Number of \qpredslong.}
The more predicates in the query, the more facts in the knowledge base have to be traversed to resolve the query and, hence, the more time it takes for the query to complete. For instance, among two non-recursive queries, Q1 and Q6, with one and two \qpredslong, the former takes less time ($0.59 \mu s$) than the latter ($2.5 \mu s$) to complete when executed on the facts obtained from a chain of sixteen NFs. 
Similarly, Q3 with two \qpredslong finishes faster ($1.4 \mu s$) than Q4 and Q5, which have three \qpredslong ($1.6 \mu s$ and $3.3 \mu s$). We also observed that the combination queries with more than three \qpredslong (Q(3,4), Q(4,5), Q(3,5), Q(3,4,5) and Q(6,7,8)) took a longer time ($6.6 \mu s$, $8.9 \mu s$, $4.7 \mu s$ and $9.6 \mu s$) comparatively to complete than the queries with three or less \qpredslong (Q(1,2) and Q(7,8)) which took comparatively less time ($1.48 \mu s$ and $2.25 \mu s$) when executed on the facts corresponding to the chain of sixteen NFs. Additionally, the time grows linearly with the \#\qpredslong.

\myparab{NF-Chain length.}
We observe that the number of instructions to be analyzed increases as the NF chain length increases, resulting in a large set of facts being generated for each NF in the chain. Consequently, in general, the query execution time of both individual and combination queries increases\footnote{There can be some exceptions due to the optimizations on indexing and caching mechanism of the Prolog runtime engine.}. In other words, the longer the NF chain, the more time it takes to execute the queries. 
Execution of all the $15$ queries take $27.9\mu s$, $33.2\mu s$, and $48.8\mu s$ on the facts generated by $2$, $8$, and $16$ NF-chains. 
We show the time taken by different queries and the predicates within each query on the facts generated by each of these NF-chains in \fref{fig:assertion_retrival_query_execution}.

\subsection{Memory footprint}
\label{subsec:mem_footprint}

We evaluate the memory footprint to understand the memory utilized by \oursystem\ for generating \CFG and executing queries. We use peak Resident Set Size (RSS) in MBs as a metric for computing memory footprint. The peak RSS is the maximum memory consumed by the process during its lifetime. We use the GNU \texttt{time} command~\cite{time_command} to measure the peak RSS value.
We compute memory footprint while (i) generating \CFG of the microbenchmark NFs listed in Table~\ref{tab:microbenchmark}, and (ii) executing individual assertion queries (Q1--Q8) and their combinations listed in Table~\ref{tab:queries-eval}. 

% \vspace{0.2cm}
\myparab{Peak RSS for \CFG generation.} 
We show the peak RSS values obtained while generating the \CFG in \fref{fig:time_rss_gen} (right). We analyzed the correlation of peak RSS with the number of paths in the Control Flow Graph (CFG) and the number of instructions analyzed (NAI). We observe that memory usage increases with the number of paths and instructions. Importantly, \oursystem\ consumes less than 50 MB for the largest microbenchmark NFs (NF13--NF16).

% \vspace{0.2cm}
\myparab{Peak RSS for Query execution.}
We observe that the peak RSS size remains almost constant for answering different queries for a given set of facts. Specifically, for the facts generated from all the 16 NFs, the Query Engine consumes $13.25$--$13.42$ MB of memory for answering our queries.

\myparab{Comparison with Klint.} \fref{fig:Yaksha_klint_verification_time} (right) shows peak RSS memory usage during verification. Klint requires $190$–$1000$ MB, while \archp uses only $20$-$50$ MB. Despite storing the network context of the entire bytecode, \archp consumed $90$-$95$\% less memory than Klint.

\subsection{Semantic equivalence of rewritten eBPF programs for comparison with Klint}
\label{subsec:semantic_equivalence_of_rewritten_eBPF_programs}
To ensure semantic equivalence in our comparison with Klint, we followed a three-step methodology.

\textbf{Logic-preserving rewriting.}
We manually rewrote Klint’s C-based NFs as XDP-compatible eBPF programs. During this process, we carefully preserved the original packet-processing logic, control flow, and state updates. The rewritten programs were manually inspected to confirm that they implement equivalent packet-processing behavior.

\textbf{Equivalent simplification.}
In cases where eBPF constraints prevented direct translation of certain components (e.g., complex hash functions), we applied equivalent simplifications to both implementations. For example, we replaced complex hash functions with simpler alternatives in both versions to maintain comparable computational complexity while preserving functional intent.

\textbf{Configuration alignment.}
We aligned configuration parameters across implementations, including the number of interfaces and table/map sizes, to avoid discrepancies arising from differing resource allocations.

We note that program equivalence is \textit{undecidable} in general. Our validation therefore relies on careful manual inspection and controlled alignment rather than formal equivalence checking.

\end{document}